\documentclass[fleqn,12pt]{wlscirep}
\usepackage[utf8]{inputenc}
\usepackage[T1]{fontenc}
\usepackage{float}
\usepackage{subfigure}
\title{Untangling Climate's Complexity: Methodological Insights}

\author[1,]{Alka Yadav}
\author[2,*]{Sourish Das}
\author[1,**]{Anirban Chakraborti}

\affil[1]{School of Computational and Integrative Sciences, Jawaharlal Nehru University, New Delhi-110067, India}
\affil[2]{Chennai Mathematical Institute, Chennai-603103, Tamil Nadu, India}
\affil[*]{sourish@cmi.ac.in}
\affil[**]{anirban@jnu.ac.in }


\begin{abstract}

In this article, we review the interdisciplinary techniques (borrowed from physics, mathematics, statistics, machine-learning, etc.) and methodological framework that we have used to understand climate systems, which serve as examples of ``complex systems''. We believe that this would offer valuable insights to comprehend the complexity of climate variability and pave the way for drafting policies for action against climate change, etc.
Our basic aim is to analyse time-series data structures across diverse climate parameters, extract Fourier-transformed features to recognize and model the trends/seasonalities in the climate variables using standard methods like detrended residual series analyses, correlation structures among climate parameters, Granger causal models, and other statistical machine-learning techniques. We cite and briefly explain two case studies: (i) the relationship between the Standardised Precipitation Index (SPI) and specific climate variables including Sea Surface Temperature (SST), El Niño Southern Oscillation (ENSO), and Indian Ocean Dipole (IOD), uncovering temporal shifts in correlations between SPI and these variables, and reveal complex patterns that drive drought and wet climate conditions in South-West Australia; (ii) the complex interactions of North Atlantic Oscillation (NAO) index, with SST and sea ice extent (SIE), potentially arising from positive feedback loops.

\end{abstract}
\begin{document}

\flushbottom
\maketitle
\section{Introduction}

\begin{quote}
{\it ``If you think the economy is more important than the environment, try holding your breath while counting your money.''} --
Professor Guy McPherson
\end{quote}

The climate system is an excellent example of a ``complex system'', which is composed of many interconnected and interdependent parts that exhibit ``emergent'' behaviors and properties not easily predictable from the behavior of individual parts or ``sum of its parts'' \cite{parisi1999, vemuri2014modeling, gell2002complexity}. Complex systems are ubiquitous and hence are studied in various domains such as physics, biology, ecology, sociology, economics, environmental science, etc. They exhibit characteristics like: (i) non-linearity: Small changes in one part of the system can lead to significant and often unpredictable effects throughout the system; (ii) emergence: Novel properties or behaviors emerge at higher levels of organization that are not directly attributable to the individual components of the system; (iii) dynamical behavior and adaptation: systems often exhibit dynamic behaviors such as self-organization, chaos, phase transitions, and they have the ability to adapt and evolve in response to changes in their environment or internal dynamics; (iv) feedback loops: Interactions among system components create feedback loops, where the output of a process feeds back into the system, influencing further interactions, and often leading to catastrophic instabilities \cite{chakraborti2020emerging}.
Understanding the dynamics of complex systems, therefore, necessitates a multidisciplinary approach integrating mathematics, physics, statistics, machine learning, and other tools of data science \cite{chakrabarti2023data}. We must mention here the work of Klaus Hasselmann, a German climate scientist and Nobel Laureate, who made substantial contributions to our understanding of climate dynamics and the development of climate models. In his groundbreaking 1976 work \cite{Hasselmann1976}, Hasselmann introduced the concept of stochastic climate modeling, incorporating random processes into climate models to elucidate how natural variability and random factors can influence long-term climate trends. He also devised statistical methods to distinguish between natural climate variability and human-induced effects, providing strong evidence of human impact on global warming. Additionally, he played a key role in developing coupled atmosphere-ocean models, crucial for accurately simulating and predicting climate dynamics. In 2021, Hasselmann, alongside Syukuro Manabe and Giorgio Parisi, received the Nobel Prize in Physics \cite{nobel2021} for his pioneering contributions to climate modeling and the understanding of complex physical systems.
In many social and environmental systems (including climate), we often do not have a clear understanding of the causal relationships between different variables. This makes the understanding of complex climate dynamics significantly more  challenging. Therefore, we take the help of Granger causal models \cite{Granger1969} and other tools of statistical inference \cite{Tibshirani1996, Tibshirani2013} to deepen our comprehension of the dynamic interactions among key climate variables, and expand our insights into the intricate mechanisms shaping climate patterns.  

Climate change poses a significant challenge globally, affecting ecosystems, economies, well-being of humans as well as livestock \cite{unclimatechange, waclimatechange, waclimatetrends, waclimate}. 
\begin{quote}
{\it ``Climate change is the greatest threat we face. It's the defining issue of our time, and we have to address it if we want to leave a thriving planet for future generations.''} --
Katharine Hayhoe, a climate scientist and professor
\end{quote}
We hope that understanding of the complex processes behind phenomena of climate change, etc. using insights from multiple disciplines will help us develop effective strategies for mitigation and adaptation. In this respect, we must mention the pioneering works of the economist and Nobel Laureate, William Nordhaus, who developed dynamic and quantitative models (now called integrated assessment models) that described the global interplay between society, the economy, and climate change \cite{Nordhaus1993}. The present interdisciplinary research approaches may supplement existing models and their applications in public policy (see e.g., Epilogue in Chakraborti et al. \cite{chakraborti2023quantum}). 

Traditionally, people have been building climate models with  computer-based simulations of mathematical equations to represent the interactions and processes within the earth's climate system \cite{schneiderclimate, hoskinsclimate, isaacClimateModeling}. These models have been then used to understand the past climate variations, predict future climate trends, and also assess the potential impacts of climate change. Generally, climate models integrate data on atmospheric dynamics (simulations of the movement of air masses, circulation patterns, and atmospheric processes such as convection, precipitation, and radiation), ocean dynamics (currents, temperature variations, and interactions between the ocean and atmosphere, including El Niño and La Niña), land surface properties (vegetation cover, soil moisture, and land use changes, which influence energy and water exchanges with the atmosphere), and other relevant factors like carbon, nitrogen, and other biogeochemical cycles that play a role in regulating the Earth's climate. Hence, the computer-based climate models provide a physical foundation for the climate change projections and are therefore built to include some of the most comprehensive range of physical, chemical, and biological processes with immense computational complexity; thereby calling for simpler approaches \cite{polvaniclimate}. In the recent past, we too have proposed alternate and simple approaches, based on machine learning and data science, to identify significant statistical relationships among these climate variables and enhance our understanding of climate dynamics \cite{YADAV2023, yadav2023NAO}.

In this review article, the basic aim is to introduce the simple methods we have used to analyse time-series data structures across diverse climate parameters, extract Fourier-transformed features to recognize and model the trends/seasonalities in the climate variables. We have used standard methods like detrended residual series analyses, correlation structures among climate parameters, Granger causal models, and other statistical machine-learning techniques. Below, we explain  the background and rationale for our two case studies \cite{YADAV2023, yadav2023NAO}:
\begin{enumerate}
    \item The relationship between the Standardised Precipitation Index (SPI) and specific climate variables, including Sea Surface Temperature (SST), El Niño Southern Oscillation (ENSO), and Indian Ocean Dipole (IOD) \cite{Wenhong2008, Rengung2008, Francois2020, Wen2105, Loughran2018}, uncovering temporal shifts in correlations between SPI and these variables, and reveals complex patterns that drive drought and wet climate conditions in South-West Australia. Droughts manifest in various forms, with meteorological drought being a critical indicator of extreme climate conditions \cite{Heim}. Previous studies have explored the dynamics of meteorological droughts and their relationship with climatic factors such as ENSO and IOD cycles \cite{McKee1993, Christos2011, KEMAL2005, Francois2020, Wen2105, Loughran2018, Lee2009, Taschetto2009, Koll2011, NDEHEDEHE2021126040, zumo2014, Nicholls1989}. However, there was a necessity for multivariate approaches to understanding drought dynamics, and hence we developed Granger causal models to examine the causal relationships among the variables (SST, NINO 3.4, and IOD) and their collective impact on SPI in South-West Australia, leveraging machine-learning techniques \cite{YADAV2023}.
    \item The complex interactions of the North Atlantic Oscillation (NAO) index, with SST and sea ice extent (SIE), potentially arising from positive feedback loops. We delved into another study \cite{yadav2023NAO} focusing on the complex dynamics of climate variables such as the North Atlantic Oscillation (NAO), a key atmospheric pressure index affecting weather patterns across North America and Northern Europe. Past research has highlighted the NAO's substantial impact on cold air outbreaks, storm occurrences, and climate variability in these regions.  Previous studies had also underscored the positive feedback loop between melting Arctic SIE and increasing SST, driven by atmospheric new particle formation and growth, accelerating Arctic warming \cite{Dall2017}. Recent studies had shown: (i) a significant decrease in SIE in the coming years, intensifying global atmospheric circulation and directly impacting SIE melting \cite{Das2018, Cressey2007}. (ii) The winter NAO plays an important role in weather variability in northwest Europe, with recent studies highlighting the predictive power of autumnal Arctic sea ice for winter NAO forecasting \cite{Warner2018}. (iii) NAO variability accounts for a substantial portion of atmospheric pressure variability and correlates with SST anomalies \cite{Kwok2000, Miettinen2011, Pan2005}. Besides, climate models have illustrated how multidecadal variations in the NAO induce corresponding fluctuations in Atlantic circulation and Arctic sea ice loss, contributing to hemispheric warming \cite{Delworth2020}. Hence, we developed a hybrid model to analyse the relationships: (a) the SPI to SST, NINO 3.4, and IOD, (b) the interplay among North Atlantic Oscillation (NAO), SST, and Sea Ice Extent (SIE). Utilizing machine learning algorithms like LASSO, we identified significant Fourier harmonics essential for modeling long-term memory. Additionally, to capture short-term memory, we incorporated lagged estimators such as IOD, SST, and NINO 3.4 within the framework of a Granger causal model. Employing data-driven techniques, we revealed intricate interactions among NAO, sea surface temperature (SST), and sea ice extent (SIE), shedding light on critical instabilities and feedback loops crucial for addressing climate change \cite{Dall2017, Das2018, Cressey2007, Jaiser2012, Pan2005, Becker1996, Slonosky2001, Warner2018, Guokun2021, Horvath2021, Kwok2000, Miettinen2011, Parkinson2000, Delworth2020}. Our approach had two distinct advantages: (i) It offered a  broader perspective for addressing climate change compared to traditional climate forecast models \cite{Kolstad2019}. (ii) While previous studies had focused on specific Arctic regions and seasons, our approach employed statistical machine learning models to provide a comprehensive view of the {\it entire} North Atlantic region.
\end{enumerate}

\section{Methodology}

The primary goal is to analyse and model the climate variables and learn about the intricate interdependences. The climate variables are typically observed and analysed as multivariate time series datasets. 
Suppose, 
\begin{equation}
\mathbf{x}_t = (x_{1,t}, x_{2,t}, \ldots, x_{k,t})^T
\end{equation} 
is a vector of different climate variables observed at time point $t$; where $x_{k,t}$ represents the value of the $k^{th}$ climate variable at a time at $t$. The superscript $T$ denotes the transpose operation, which converts the row vector into a column vector.

\subsection*{Simple measures of dependences} \label{sec:acf_ccf}

\vskip 0.3cm

\noindent \textbf{Autocorrelation function}, also known as \emph{serial correlation} \cite{acf}, is a statistical concept that measures the degree of similarity between a given time series of a climate variable and a lagged version of itself over successive time intervals. It quantifies how much the current value of a climate variable is related to its past values. For example, we may be interested in knowing if the past 30 days' values of SST influence the current value of SST. If this is so, that might help us to predict the possible value of SST 30 days ahead from today. The autocorrelation function (ACF) is described as follows:
\begin{equation}
\text{ACF}(k) = \frac{\text{Cov}(x_t, x_{t-k})}{\sqrt{\text{Var}(x_t) \times \text{Var}(x_{t-k})}}
\label{eqn_acf}
\end{equation}
where ACF($k$), is the autocorrelation between the values of the climate variables at time $t$ and those at time $t-k$, and $k$ represents the number of time units (lags) by which the series is shifted. The covariance $\text{Cov}(x_t, x_{t-k})$ is between the values of the climate variables at time $t$ and those at time $t-k$; and $\text{Var}(x_t)$ and $\text{Var}(x_{t-k})$ represent the variance of the climate variables at the time $t$ and those at time $t-k$. A positive ACF value at lag $k$ indicates a positive correlation between the values of the same climate variables separated by $k$ time units. 

\vskip 0.3cm

\noindent \textbf{Cross-Correlation Function} (CCF) is a statistical measure that indicates the relationship between values of two different climate variables separated by a time lag \cite{ccf}. 
The cross-correlation function (CCF) is defined as:
\begin{equation}
   \text{CCF}(k) = \frac{\text{Cov}(x_t, y_{t-k})}{\sqrt{\text{Var}(x_t) \times \text{Var}(y_{t-k})}} 
   \label{eqn_ccf}
\end{equation}
where $\text{Cov}(x_t, y_{t-k})$ is the covariance between the variable $x_t$ at time t and the lagged value of variable $y_t$ at time $t-k$. The CCF can take on positive, negative, or zero values. A positive CCF indicates a positive correlation between the two variables at the specified lag, while a negative CCF indicates a negative correlation. A CCF close to zero suggests no significant correlation between the variables at that lag. 

\subsubsection*{Calculating Periodicity using ACF} \label{sec:periodicity}

For examining the enduring memory span of each time series, which indicates the persistence of past observations influencing current values, we utilised the following algorithms:

\begin{enumerate}
\item[(i)] Calculating the ACF (Eqn. \ref{eqn_acf}) on training data.
\item[(ii)] Identifying periods $P_1, P_2,\cdots, P_{m_s}$, where the autocorrelation $\rho_m$ exceeds a threshold $s$, with $s=\frac{1}{M}\sum_{m=1}^M|\rho_m-\hat{\rho}|$; here, $\rho_m$ represents the $m^{th}$ lag autocorrelation, $\hat{\rho}$ is the median of all autocorrelations, and $M$ represents the maximum lag considered in our invastigation study. 
\end{enumerate}

To illustrate this algorithm, we take the example of the SPI time series of West South Australia (coordinates: longitude 113.7158 and latitude -26.6969) and plot the ACF (shown later). Utilising a dataset spanning 450 months, ranging from June 1973 to November 2010, we detected three noteworthy periods: 216, 151, and 60 months. These periods suggested that the current SPI value exhibited a notable positive correlation with past SPI values, with periodicities of approximately 5.5 years for the specified location.

\subsection*{Modelling Temporal Structure and Seasonalities}\label{sec: Modl_temporal}

Temporal patterns in data can be effectively modeled to uncover underlying trends and periodic behavior. We introduce a comprehensive model ($M_s$) to analyse the components contributing to the observed variability:
\begin{eqnarray}
\label{eqn_full_model} M_s:~~~y(t)=\beta_0+ \beta_1 t+ \eta(t) +z(t),
\end{eqnarray}
where ($M_s$) is the model for a location $s$ time series, $y(t)$ represents the observed variable at time $t$, $\beta_0$ denotes the intercept, $\beta_1$ is the trend coefficient, $\eta(t)$ accounts for the periodicity of the process, and $z(t)$ captures short-term memory using the Granger Causal Model (explained below).

 In time series analysis, capturing seasonality is important for understanding recurring patterns. Here, we present a model ($\eta(t)$) aimed at quantifying seasonal effects:
 
    \begin{eqnarray}
    \label{eqn_model1}
      \eta(t)=\sum_{j=1}^{m_s}&\bigg\{&\sum_{i=1}^K \beta_{ji}\sin(i*\omega_j*t)  +\sum_{i=1}^{K}\gamma_{ji}\cos(i*\omega_j*t)\bigg\}.
    \end{eqnarray}
This equation breaks down seasonal variation into Fourier series, where $\omega_j=\frac{2\pi}{P_j},~~ j =1,2,\cdots,m_s$, $P_j$ is estimated via ACF, and $m_s$ denotes the number of periods for $s^{th}$ location time series. Then, we typically applied the LASSO (least absolute shrinkage and selection operator) technique\cite{LASSO1996}, a machine learning shrinkage method, to identify the most significant harmonics in the Fourier model. LASSO selects harmonics that significantly reduce error, enhancing model accuracy. 

\subsection*{Modelling Spatial Correlation} \label{sec:model_spatial}
\noindent To estimate $y(t)$ at location $s$ while accounting for spatial correlation, we introduce the estimated value $\hat{y}_s(t)$ for the $t^{th}$ month, employing a spatially correlated Gaussian process model, 

    \begin{equation}
    \bar{y}_s(t)=\Sigma(s,s')[\Sigma(s,s')+\tau^2\mathbf{I}]^{-1}\hat{y}(t),
    \end{equation}
 where, ${y}_s(t)$follows a Gaussian process with a mean of zero and a covariance function defined as $\Sigma(s,s')=\tau^2\exp{-\rho|s-s'|^2}$.
Note that the covariance matrix $\Sigma(s,s')$ models the spatial correlation between the location $s$ and $s'$ for which $\mathbb{V}ar(\epsilon)=\sigma^2$, and $\Sigma(s,s')=exp\{-\rho|s-s|^2\}$. 

\subsection*{Granger-causal model}\label{sec:Granger_Causal}
For the short-term memory or autoregressive structure, denoted as $z(t)$ (mentioned above), we employ the Granger causal model.
The Granger causality model serves as a powerful tool for assessing the changing causal dynamics between variables over time \cite{Granger1969}. Here, we present hypotheses regarding the influence of one variable on another using this framework.\\

\noindent Null Hypothesis (\textit{$H_0$}):

The null hypothesis represents that the variable of interest ($Z$) is solely dependent on its own historical memory, without any influence from other variables. It can be represented as:

\begin{equation}
\label{null_eqn_granger}
	z(t)=\beta_0+\beta_1 z(t-1)+\cdots+\beta_k z(t-k)+\epsilon(t).
\end{equation}
This equation represents a time series model where \( z(t) \) is the variable of interest at time \( t \). The term \( \beta_0 \) is the intercept, and \( \beta_1, \beta_2, \ldots, \beta_k \) are coefficients that indicate the influence of past values of \( z(t) \) on its current value. The \( \epsilon(t) \) term represents the error or residual component of the model at time \( t \), assumed to follow a normal distribution with mean 0 and variance \( \sigma^2 \).\\
	
\noindent Alternate Hypothesis (\textit{$H_a$}): 

Under the alternate hypothesis, we propose that the variable of interest ($ z(t) $) is not only influenced by its own historical memory but also by the past values of another variable ($ x(t) $). This can be represented as follows:

\begin{eqnarray}
\label{Alt_modl_SIE_res}
z(t) &=& \beta_0 + \beta_1 z(t-1) + \cdots + \beta_k z(t-k) \nonumber \\
&& + \gamma_1 x(t-1) + \cdots + \gamma_k x(t-k) + \epsilon(t). 
\end{eqnarray}
In this equation, the coefficients \( \gamma_1, \gamma_2, \ldots, \gamma_k \) represent the influence of past values of the variable \( x(t) \) on the current value of \( z(t) \). Each \( \gamma_i \) indicates the strength and direction of influence from the corresponding lagged value of \( x(t) \). This model allows us to investigate whether the variable of interest ($ z(t) $) is affected by the past values of another variable ($ x(t) $), thereby examining potential causal relationships between them.

To determine whether the null hypothesis should be rejected, we assess whether all coefficients \( \gamma_i \) in the alternate hypothesis are equal to zero. Specifically, the null hypothesis \( H_0 \) states that \( \gamma_1 = \cdots = \gamma_k = 0 \). To reject this null hypothesis in favor of our alternate hypothesis \( H_a \), we need to ascertain if at least one coefficient \( \gamma_i \) is not equal to zero. This rejection allows us to clearly understand the effect of the variables under consideration.


\section{Two case studies of climate complexity}
\subsection{Complex Dynamics of Drought in South-West Australia}

Our first study delves into the complex interactions among climatic variables such as SST, NINO 3.4, and IOD, examining their influence on the SPI in South-West Australia.

\subsubsection*{Data Description}

We utilised four primary climate variables: SPI, SST, NINO 3.4, and IOD indices, with a focus on South-West Australia. Our analysis involved SPI monthly time series data spanning 58 years (1961–2018), obtained from daily precipitation observations across 194 stations in the region, sourced from the Bureau of Meteorology (BOM) website \cite{footnotet4}. The datasets for NINO 3.4 and IOD were sourced from the NOAA website \cite{footnotet3, SST_Data_Source}. Our study area encompasses longitudes 113.72 to 137.12 and latitudes -26.70 to -35.73. Analyses were conducted using monthly averaged SST data from 1982 to 2018 \cite{YADAV2023}.

\begin{figure}[H]
    \centering
    \subfigure[]{\includegraphics[width=0.32\textwidth]{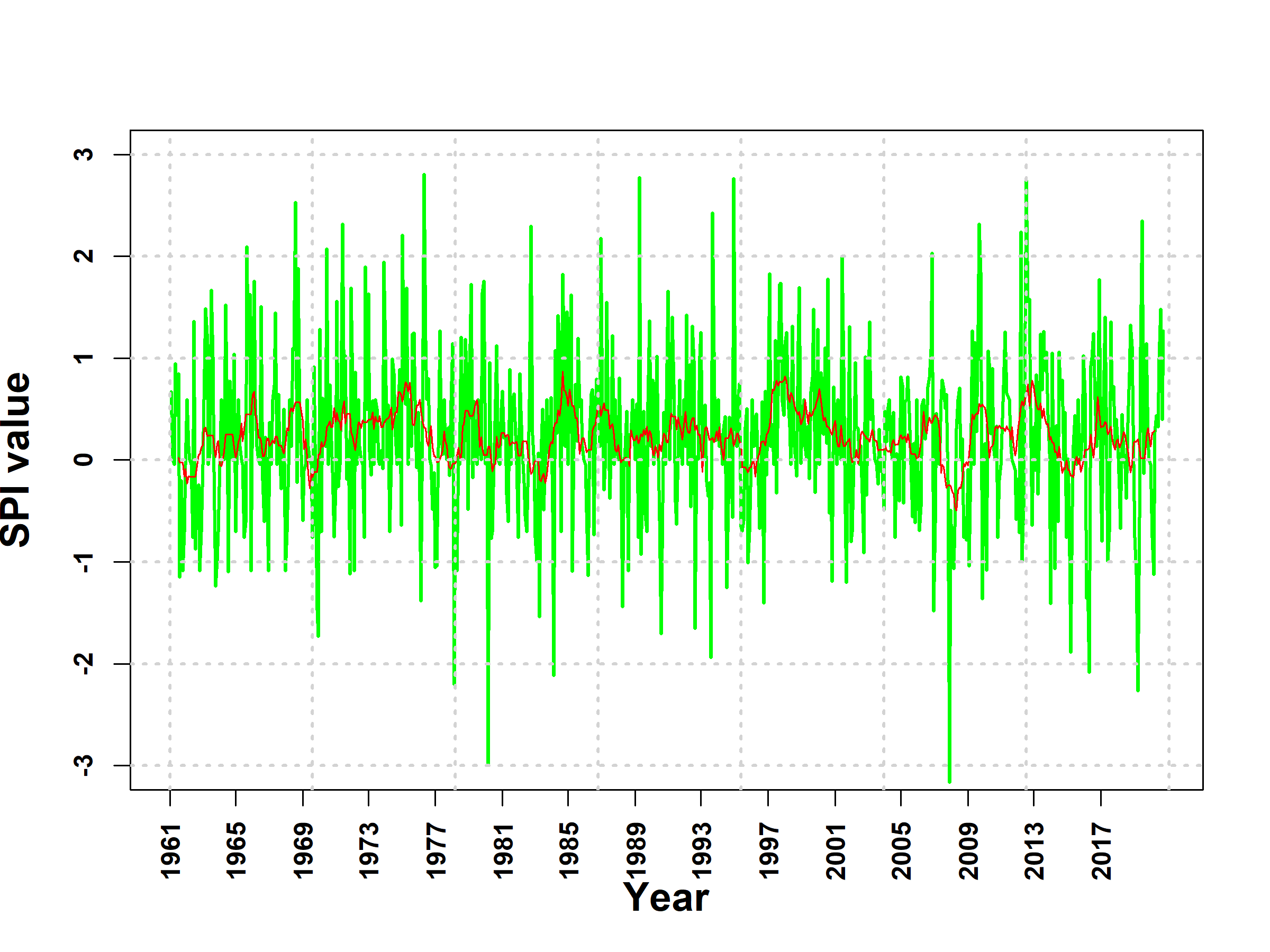}}
    \subfigure[]{\includegraphics[width=0.32\textwidth]{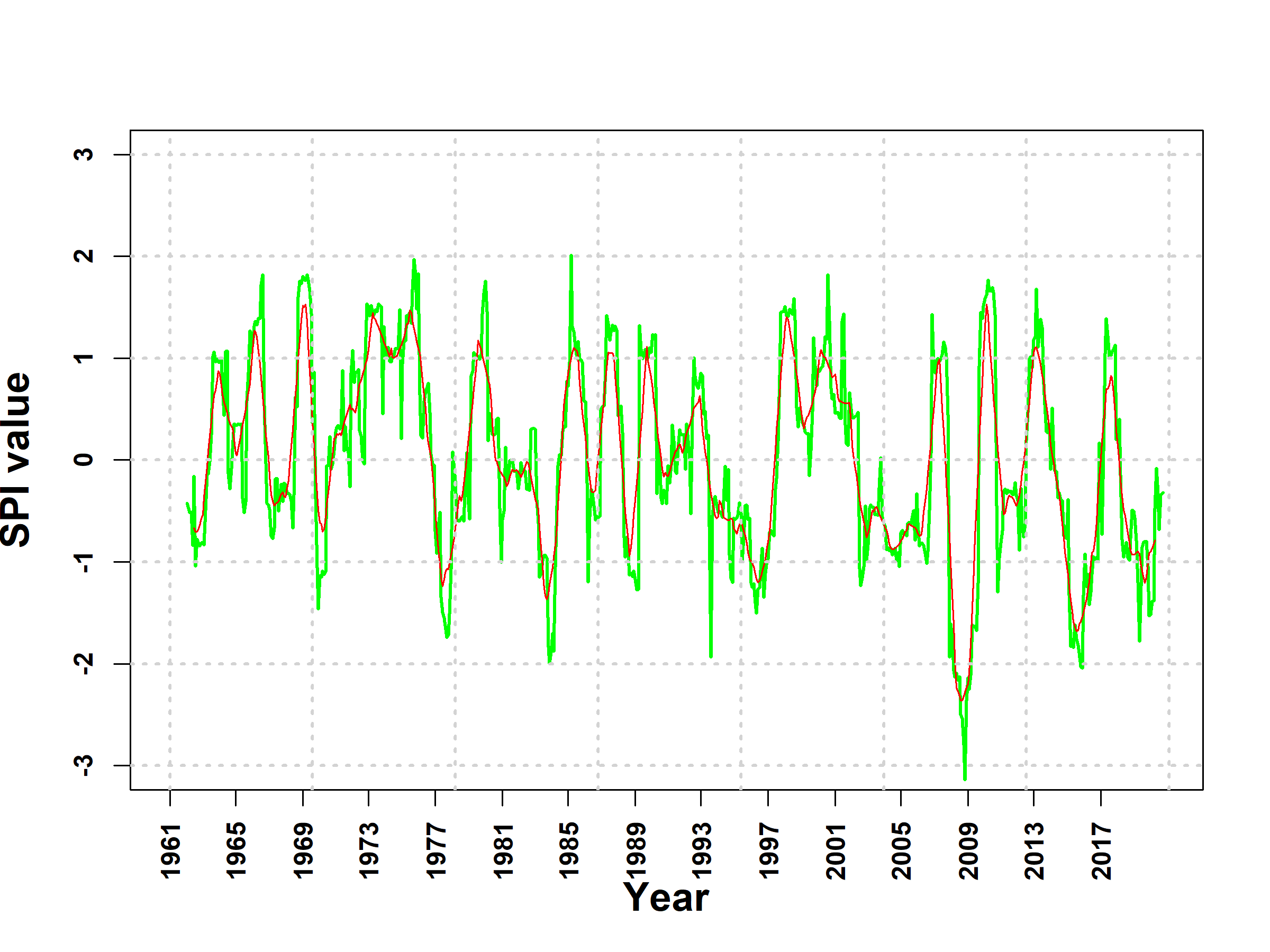}}
  
    \caption{Time series plots illustrating the Standardised Precipitation Index (SPI) for the south-western region of Australia are depicted for (a) 1-month and (b) 12-month periods, covering a span of 58 years (1961-2018). In both plots, the 12-month moving average is represented by the solid red line. A notable increase in the signal-to-noise ratio is observed in the SPI 12-month series compared to the SPI 1-month series, thereby enhancing the reliability of our analysis \cite{Lehner2006}. This consistent observation was evident across all 194 monitoring stations, prompting our selection of the SPI 12-month series for further analysis.}
    
    \label{fig_ts_spi}
    \end{figure}

\begin{figure}[H]
    \centering
    \subfigure[]{\includegraphics[width=0.32\textwidth]{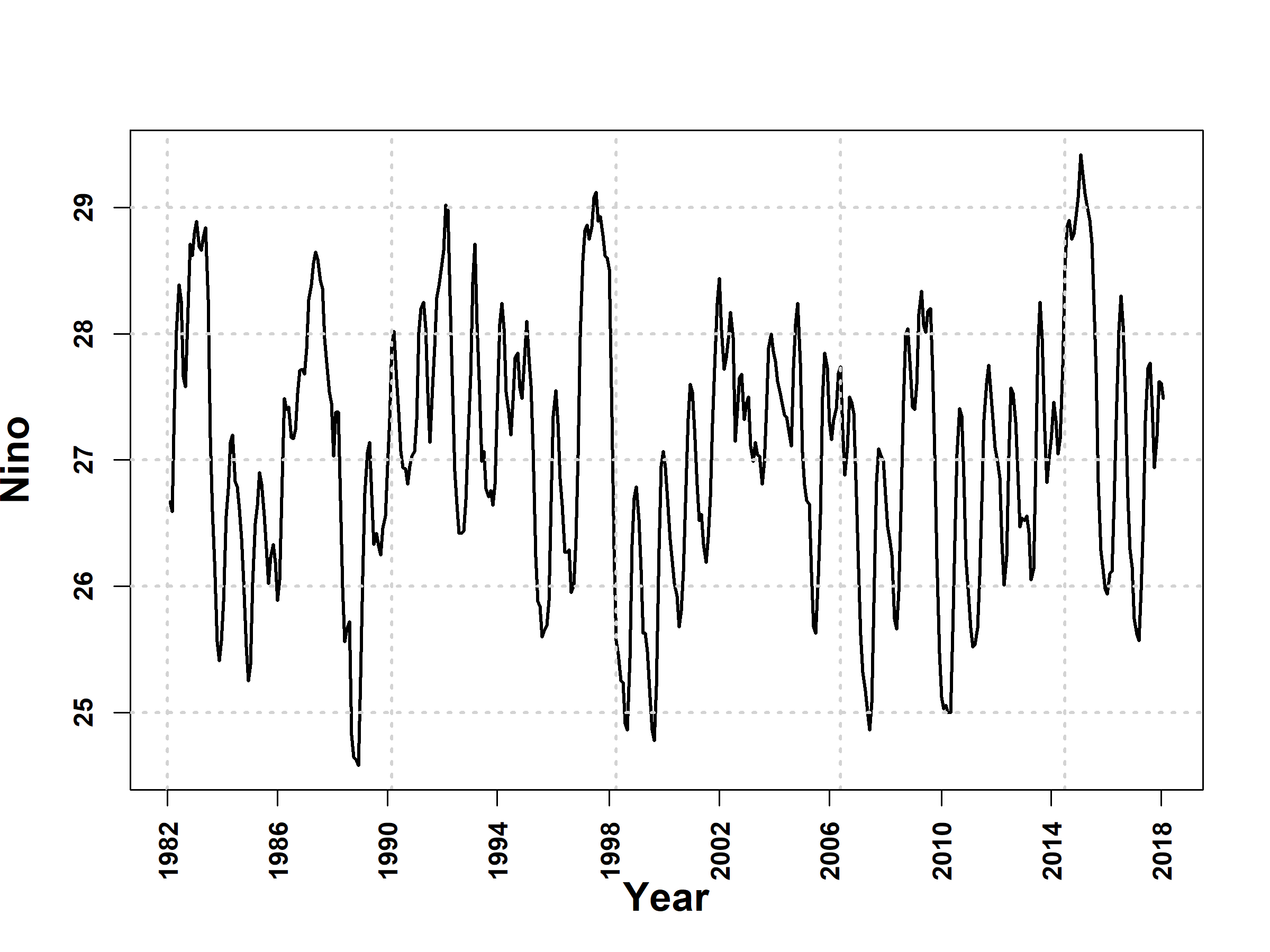}}
    \subfigure[]{\includegraphics[width=0.32\textwidth]{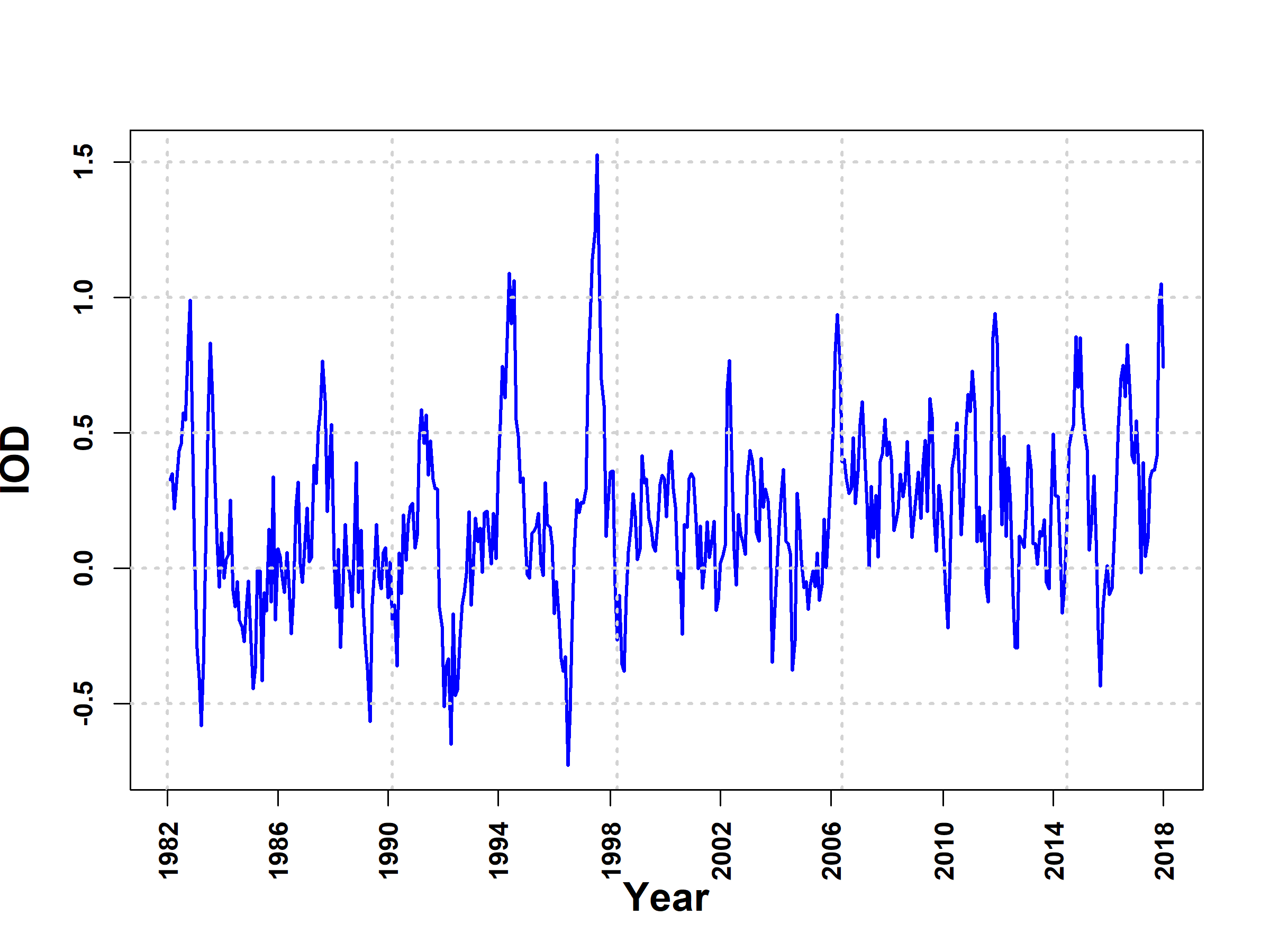}}
     \subfigure[]{\includegraphics[width=0.32\textwidth]{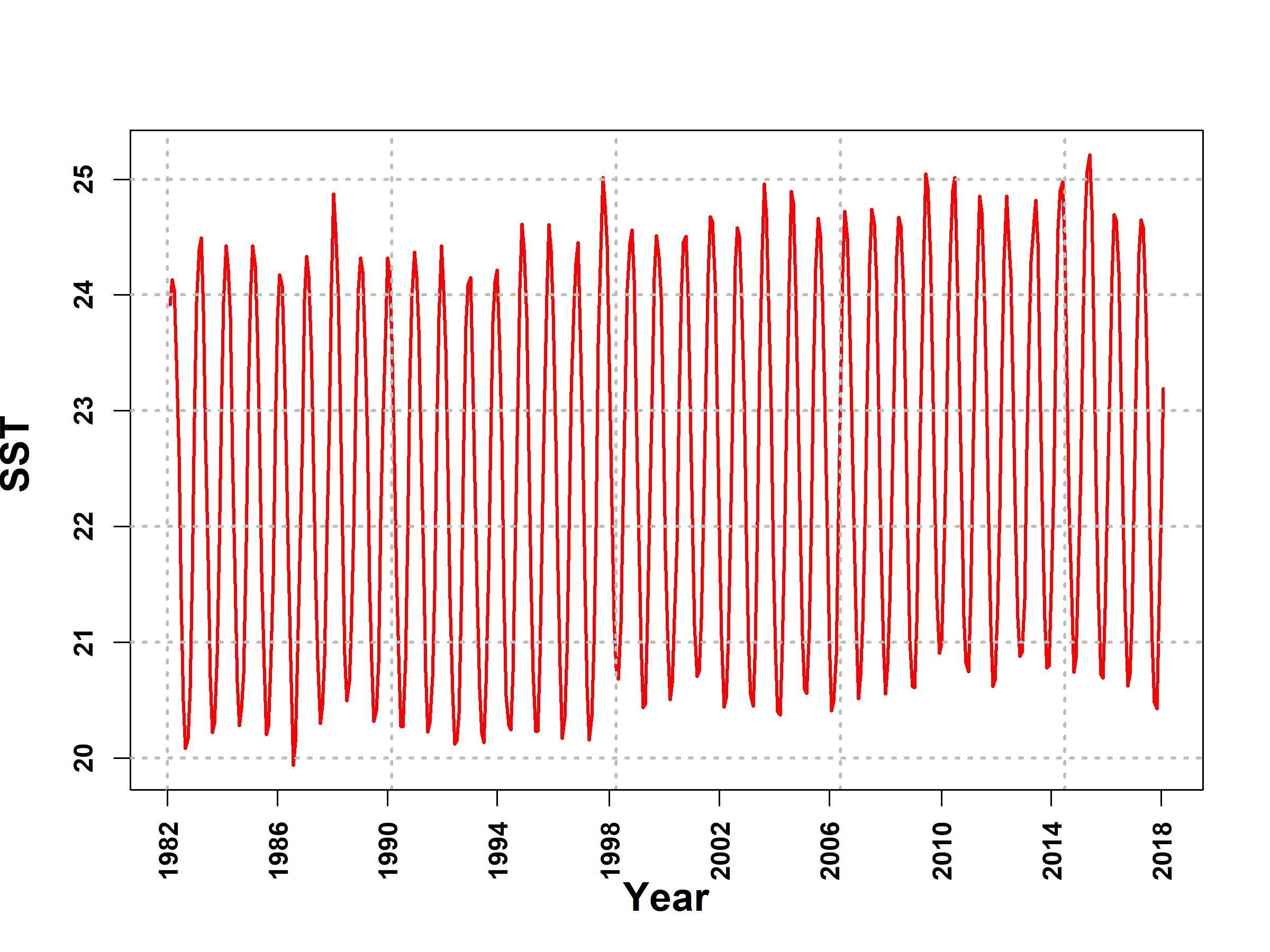}}
  
    \caption{Time-series plots illustrating (a) El Niño–Southern Oscillation (ENSO) NINO 3.4, (b) Indian Ocean Dipole (IOD), and (c) Sea Surface Temperature (SST) data are presented, covering a period of 37 years (1982-2018) for the south-western region of Australia.}
    
     \label{fig_ts_nino_iod_sst_nino}
    \end{figure}

This region holds significant agricultural importance, contributing substantially to Australia's gross agricultural production and grain exports. However, the agriculture sector faces considerable vulnerability due to the unpredictable nature of climate variables, resulting in reduced water supplies and impacting wheat and broadacre livestock production \cite{footnotet1}. Climate change exacerbates these challenges, evident from increasing temperature and fluctuating rainfall anomalies \cite{Ludwig2009,Lehner2006,waclimate,BORUFF2018,waclimatechange}. Understanding these complexities is crucial, as they severely impact agriculture, water resources, and public health. 

\subsubsection*{Results and Analyses}

Our exploration into the dynamics between the Standardised Precipitation Index (SPI) and climate variables—SST, IOD, and NINO 3.4 —across the extensive period from 1982 to 2018 revealed insightful findings. Time series plots in Figure (\ref{fig_ts_spi}) depict the SPI for both one-month and 12-month periods spanning 58 years (1961-2018). The solid red line represents the 12-month moving average, indicating a higher signal-to-noise ratio in the SPI 12-month series compared to the SPI 1-month series. This observation, supported by \cite{Lehner2006}, bolsters the reliability of our analysis. The consistency of this observation across all 194 monitoring stations prompted us to select the SPI 12-month series for our further analysis. Our investigation also unveiled a tendency for SST, Nin 3.4, and IOD to exhibit mean-reverting behavior, as evidenced in Figure (\ref{fig_ts_nino_iod_sst_nino}), where values tended to regress to their means over time. This observation aligns with findings from the Hurst exponent coefficient analysis \cite{Hurst2011},  Table (\ref{tab_Hurst_exp}) indicates a long memory characteristic in these variables, with values surpassing 0.5.

Figure (\ref{fig_acf_6045}) provides illustrative insight into one of the 194 locations studied. Utilising algorithm \ref{sec:periodicity}, we identified significant periodicities at intervals of 60, 151, and 216 months. For instance, at the specific location (longitude 113.7158 and latitude -26.6969), the SPI displayed a notable positive correlation with past SPI values, recurring at approximately 5.5-year intervals.
Such findings contribute valuable insights into the complex interplay between precipitation patterns and key climatic drivers, shedding light on the underlying mechanisms governing long-term climate variability.

 Figure (\ref{fig_Period1_194}) is a visual representation of the correlation matrix illustrating SPI among the 194 locations. This matrix underscores the spatial correlation observed across all locations, highlighting the cohesive nature of SPI dynamics throughout the region.

To further explore the dynamics, we analysed cross-correlation functions (\ref{sec:acf_ccf}) among monthly time-series of SPI, SST, IOD, and NINO 3.4 variables, as shown in Figures (\ref{fig_ccf_1999_2008}) and (\ref{fig_ccf_2009_2018}). With a maximum lag of 120 months, these CCFs reveal mutual influences among climate variables. Figure (\ref{fig_ccf_1999_2008}) illustrates the Cross-Correlation Function (CCF) spanning the years 1999 to 2008, revealing the interplay among climate variables (SST, NINO 3.4, and IOD) and their mutual influence. This interaction results in a dense network, as depicted in Figure (\ref{fig_network}a). In Figure (\ref{fig_ccf_2009_2018}), spanning 2009-2018, we observe that SST weakly couples with SPI (Figure (\ref{fig_ccf_2009_2018}a)), while IOD does not directly couple with SPI (Figure (\ref{fig_ccf_2009_2018}b)). Conversely, NINO 3.4 couples with SPI (Figure (\ref{fig_ccf_2009_2018}c)). Additionally, IOD couples with SST and NINO 3.4, though not reciprocally (Figures (\ref{fig_ccf_2009_2018}d) and (\ref{fig_ccf_2009_2018}e)). Notably, NINO 3.4 and SST exhibit mutual coupling (Figure (\ref{fig_ccf_2009_2018}f)). This comprehensive analysis, spanning 2009–2018, is depicted as a network diagram in Figure (\ref{fig_network}b).

Our hybrid model (\ref{sec: Modl_temporal}), combining Fourier harmonics for long-term memory and Granger causal modeling (\ref{sec:Granger_Causal}) for short-term memory, effectively captured SPI dynamics. Validation through Root Mean Square Error (RMSE) assessments demonstrated (Table ({\ref{tab:RMSE})) the superiority of models incorporating LASSO selection and Gaussian process (GP) spatial correction (\ref{sec:model_spatial}), particularly when considering SST as a covariate \cite{LASSO1996}. Notably, the incorporation of SST significantly improved SPI estimation accuracy.
In Figure (\ref{fig_dec_2010}), we presented the out-of-sample estimates alongside the actual SPI values. Visual examination suggests satisfactory performance of the proposed model (\ref{eqn_full_model}) for December 2010. However, it's crucial to emphasize that this validation is specific to a single month. To comprehensively evaluate the model's effectiveness, we extended the assessment period from December 2010 to November 2018 (eight years) and calculated the out-of-sample Root Mean Square Error (RMSE) for all 194 locations. The median RMSE results are outlined in Table (\ref{tab:RMSE}).
The first model (Type I) used Fourier series methods on SPI to capture long-term memory, while other models (Type II, III, IV) integrated Nino 3.4, IOD, and SST as covariates, employing various combinations and lags. Our investigation revealed that incorporating the LASSO approach for the Fourier model with spatial correction notably enhanced model performance, with Type III and Type IV models demonstrating superior efficacy. Notably, SPI 12 months exhibited lower RMSE values across all Type II, III, and IV models, indicating enhanced generalizability compared to SPI 1 month. Furthermore, our study identified SST as a significant factor influencing SPI estimation, and its inclusion as a covariate contributed to improved model performance. 

Analysis of SPI trends in Figure (\ref{fig_ts_beta_sst_nino}) revealed evolving drought conditions, with inland areas showing mild increasing trends, particularly post-2008. The negative correlation between IOD and SPI until 2008 shifted to a positive correlation thereafter, indicating a reversal in drought conditions \cite{Saji1999}. Similarly, NINO 3.4 exhibited a consistent negative relationship with SPI, suggesting wetter conditions in south-west Australia. Rising SST trends further corroborated expectations of increasingly wet conditions in the region \cite{IOD}.

Overall, our findings underscore the complex interplay of climate variables and the evolving nature of drought conditions in south-west Australia, emphasizing the importance of considering long-term trends and spatial correlations in climate modeling and prediction. Specifically, the LASSO selection process within spatial correction models, effectively enhances the accuracy of the proposed models.

\begin{figure}[H]
\centering
    \includegraphics[width=.4\linewidth]{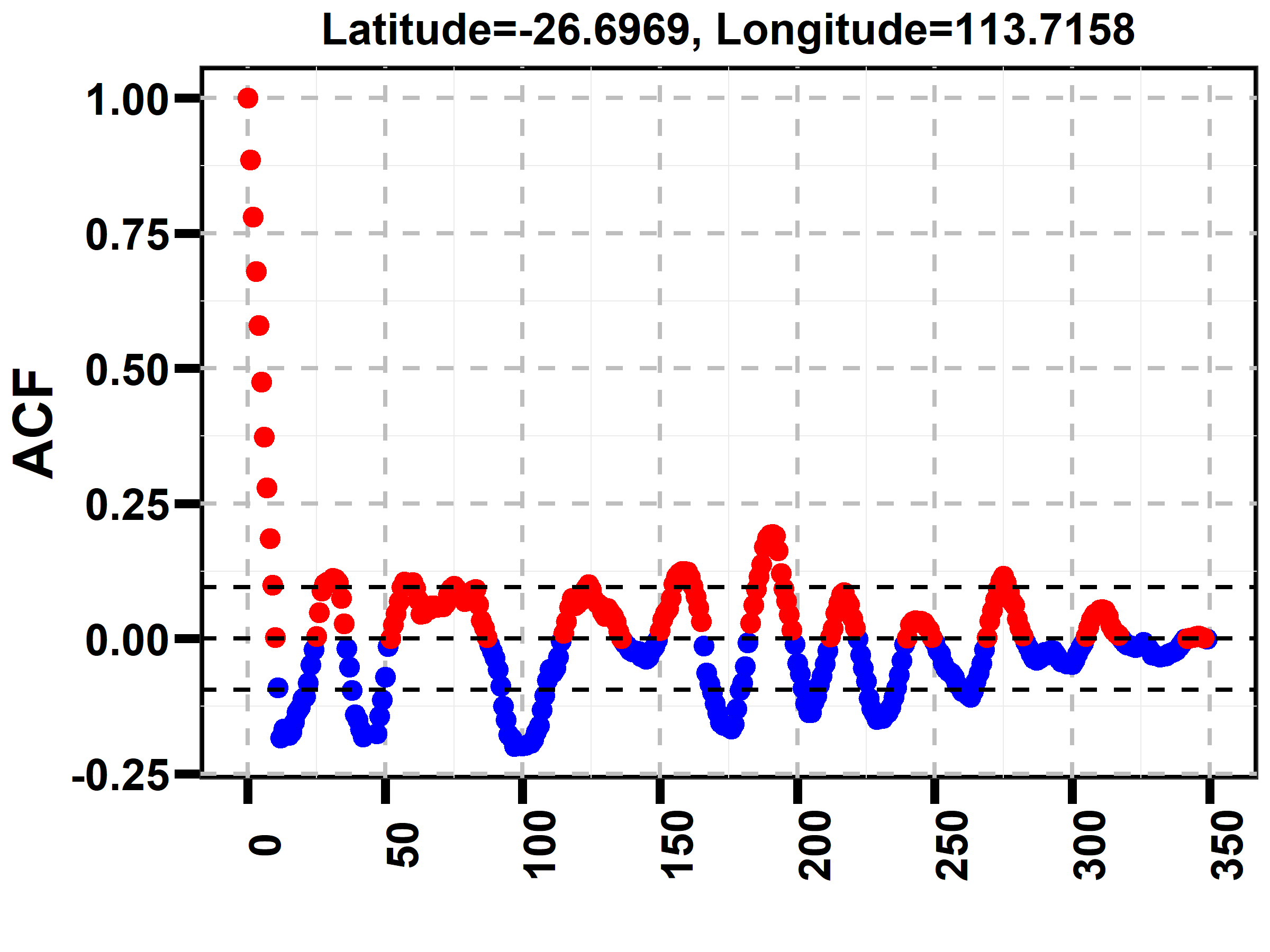}

\caption{The autocorrelation plot, with a maximum lag of 400 months, pertains to rainfall data obtained from a specific location at longitude 113.7158 and latitude -26.6969. A dataset spanning 450 months, from June 1973 to November 2010, was utilized to construct the autocorrelation function. Notably, three significant periods of 216, 151, and 60 months were identified. These periods indicate a substantial positive correlation between current and past SPI values, suggesting a periodicity of approximately 5.5 years.}
\label{fig_acf_6045}
\end{figure}

\begin{table}[H]\centering  
\setlength{\arrayrulewidth}{1mm}
\setlength{\tabcolsep}{18pt}
\renewcommand{\arraystretch}{1.5}
\begin{tabular}{p{7cm}p{3cm}} \hline
	\bf{Index}&\bf{Hurst Value} \\
	\hline
	Standard Precipitation Index (SPI)  &  0.71  ($\pm 0.05$ )\\
	El Ni\~{n}o Southern Oscillation (ENSO) NINO 3.4  & 0.66 ($\pm 0.04$)\\
	Indian Ocean Dipole (IOD)  &   0.69  ($\pm 0.06$)  \\
	Sea Surface Temperature (SST)& 0.58 ($\pm 0.05$) \\ \hline
	\end{tabular}
 \caption{This table displays the Hurst exponent values along with Bootstrap margins of error for all indices, including SPI, IOD, SST, and NINO 3.4. These values collectively suggest the presence of long memory within the system. The table was reproduced from Yadav et al. \cite{YADAV2023} }
  \label{tab_Hurst_exp}
\end{table}

\begin{figure}[ht]
\centering
\includegraphics[width=.45\textwidth]{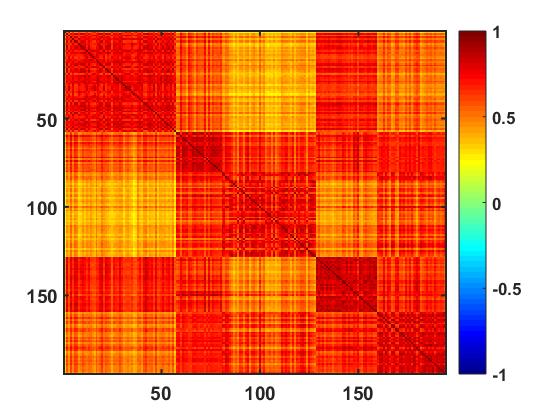}
\caption{ A visual representation of the correlation matrix depicting SPI-12 months across 194 locations reveals significant spatial correlation across all locations.}
\label{fig_Period1_194}
\end{figure}

\begin{figure}[H]

\centering
\subfigure[]{\includegraphics[width=0.32\textwidth]{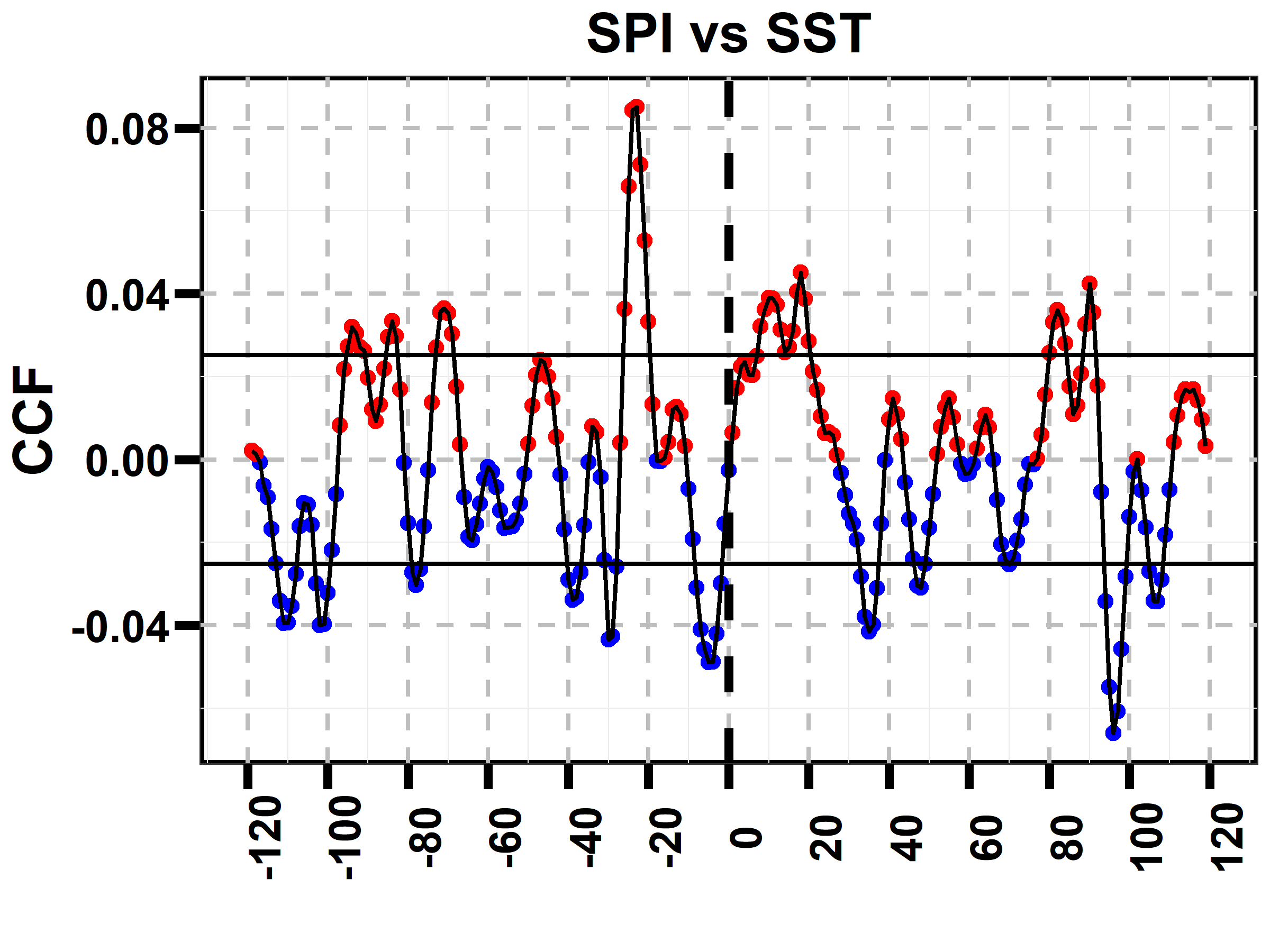}}
\subfigure[]{\includegraphics[width=0.32\textwidth]{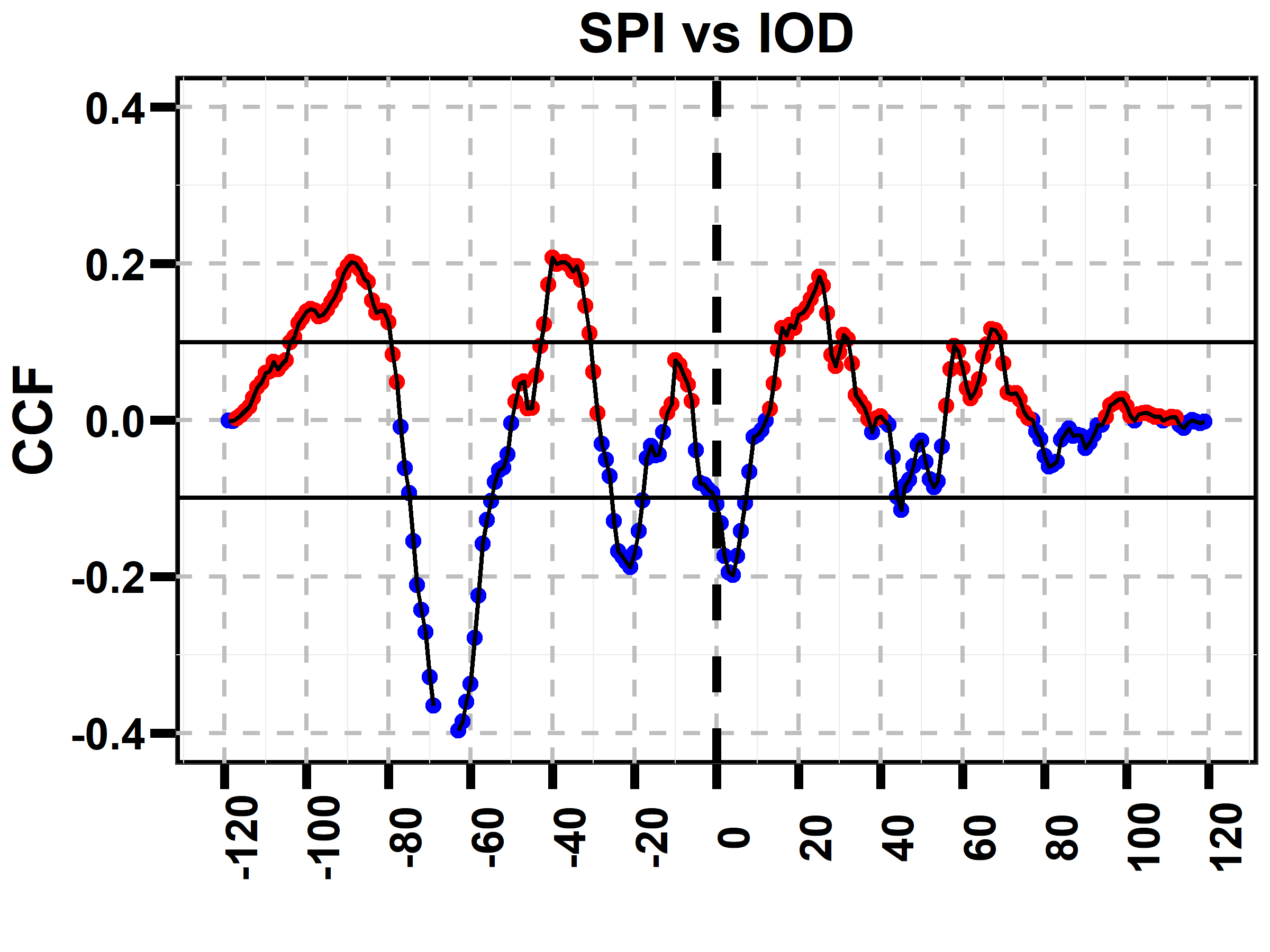}}
\subfigure[]{\includegraphics[width=0.32\textwidth]{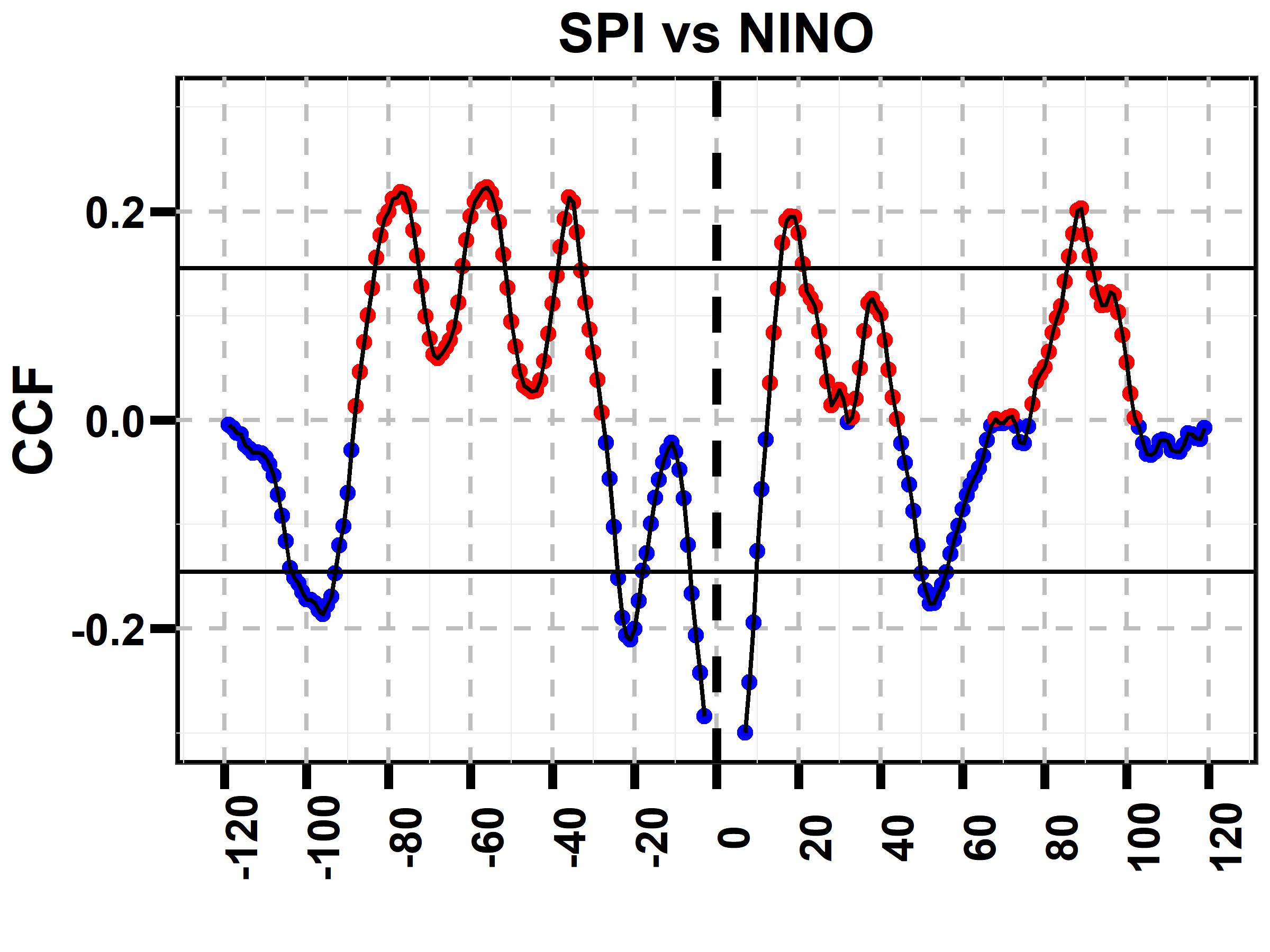}}
\subfigure[]{\includegraphics[width=0.32\textwidth]{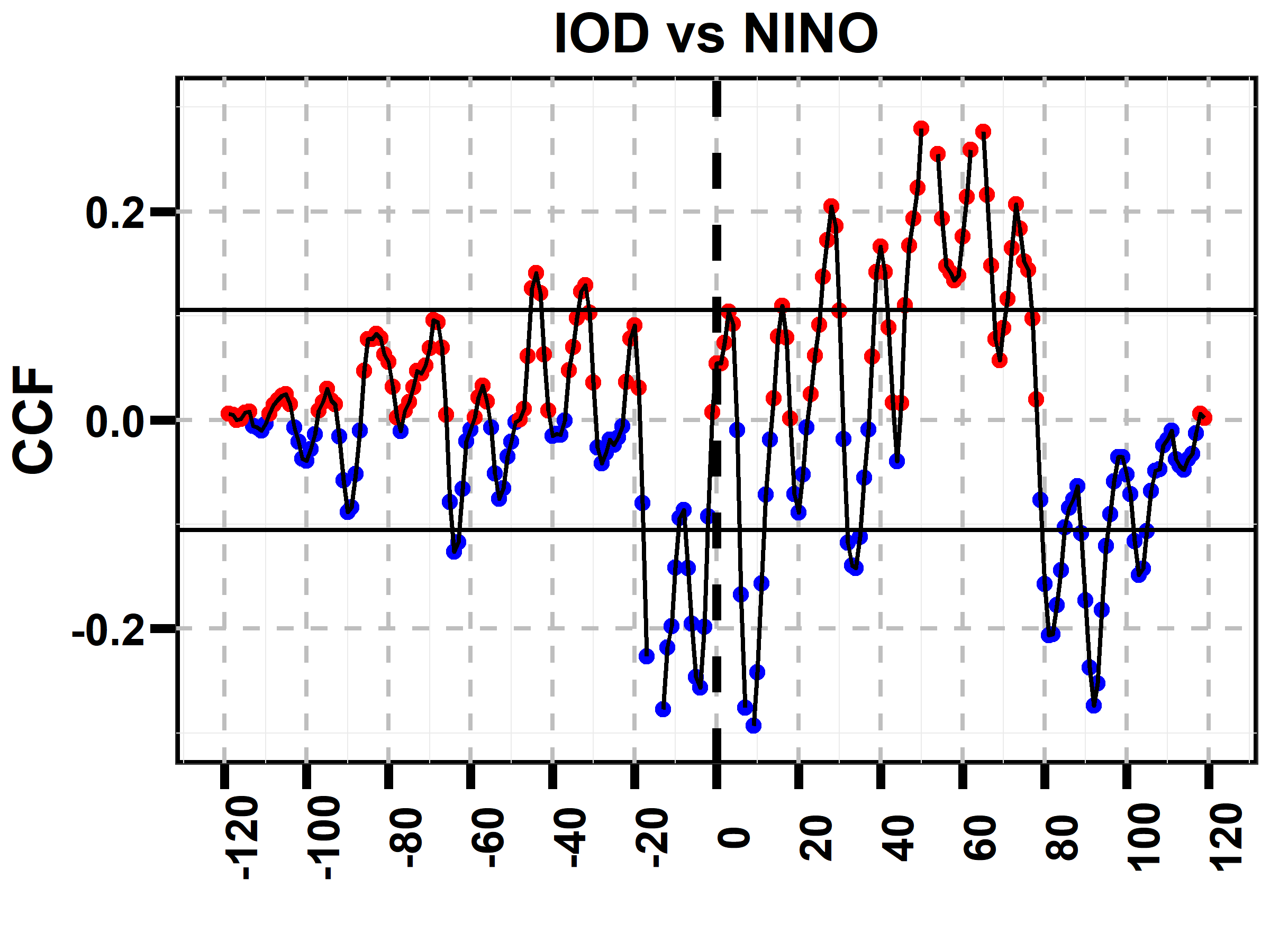}}
\subfigure[]{\includegraphics[width=0.32\textwidth]{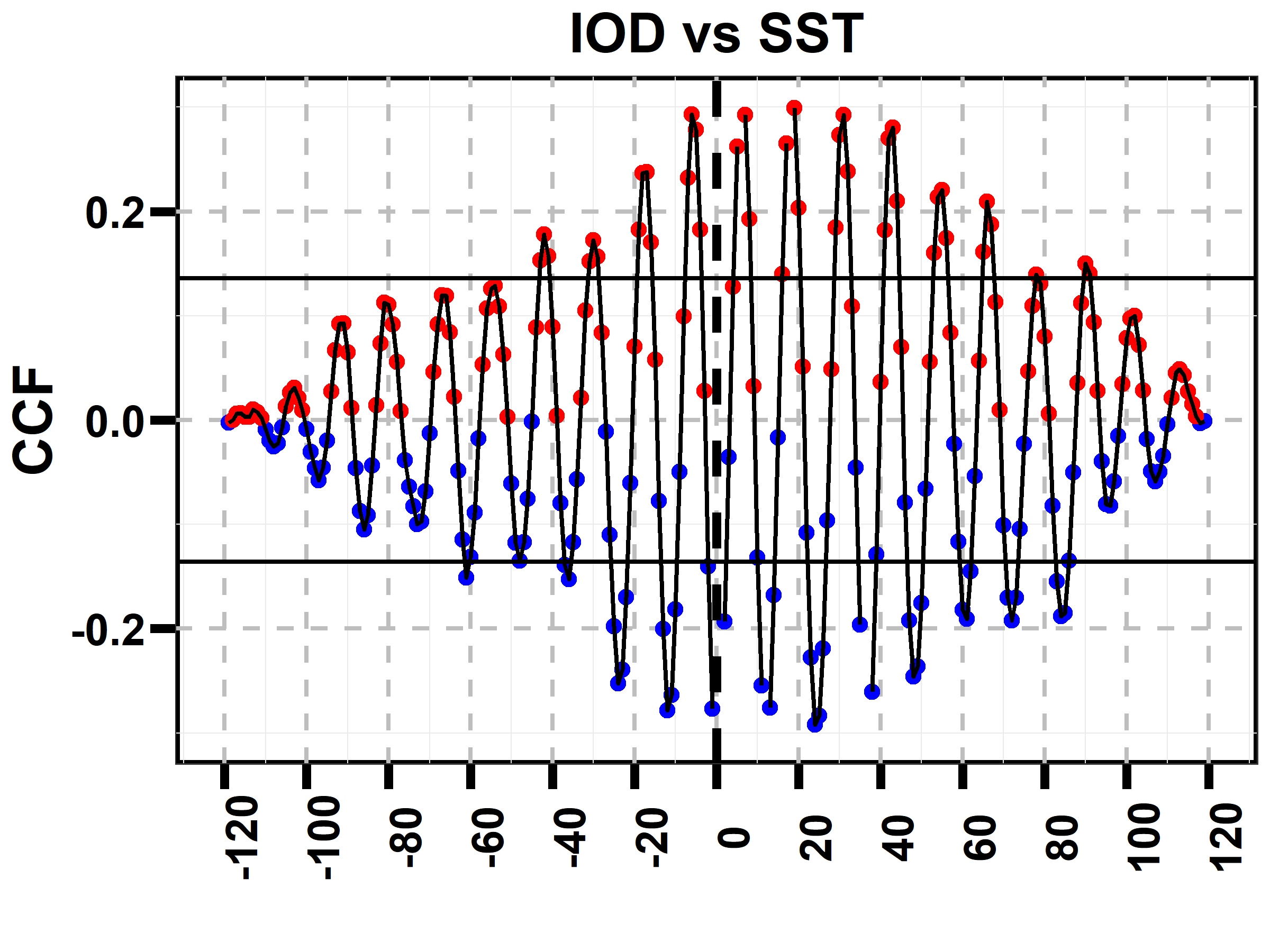}}
\subfigure[]{\includegraphics[width=0.32\textwidth]{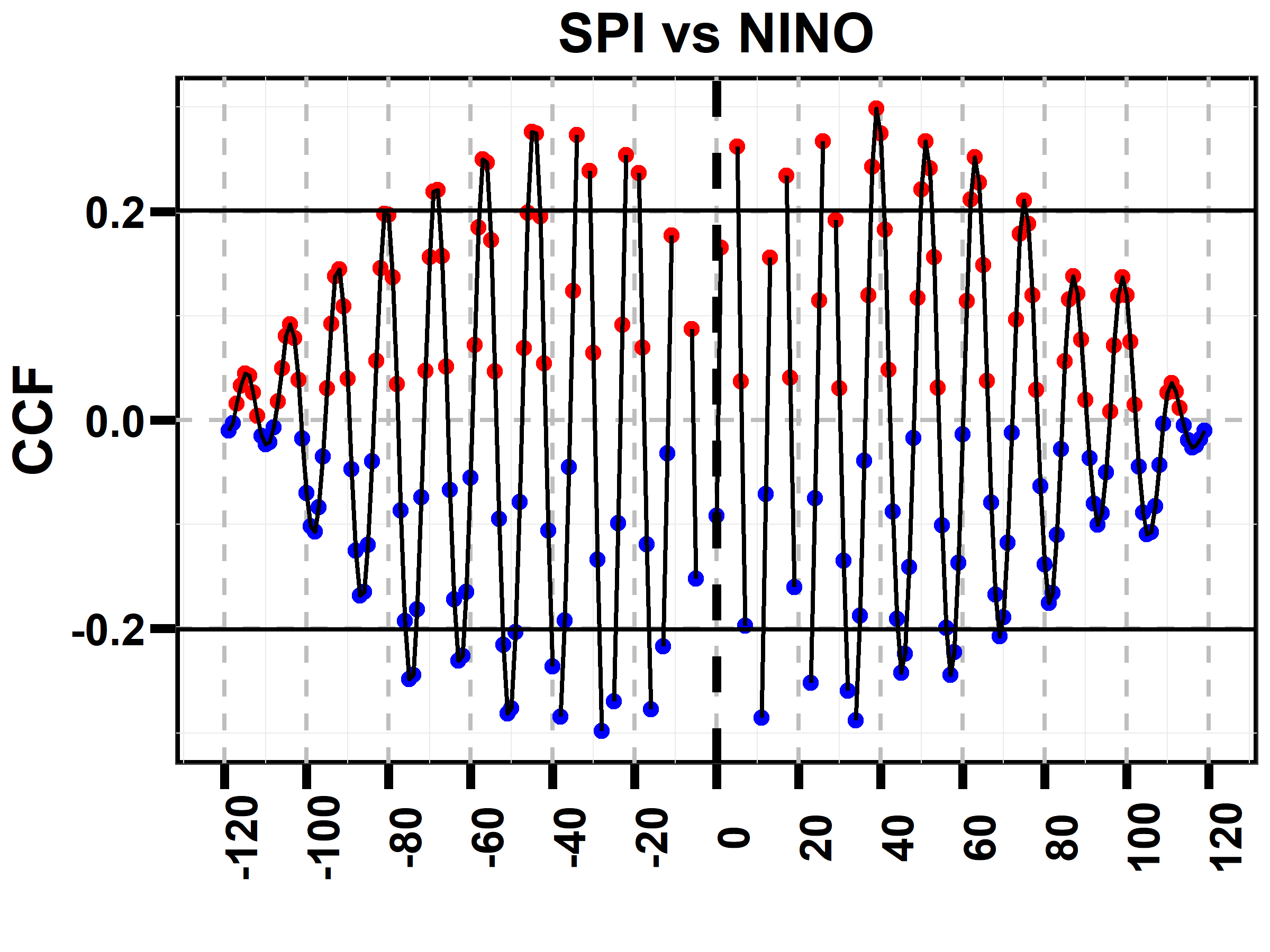}}
    \caption{During the period from 1999 to 2008, cross-correlation function (CCF) plots were generated for the following pairs: (a) Standardized Precipitation Index (SPI) and Sea Surface Temperature (SST), (b) SPI and Indian Ocean Dipole (IOD), (c) SPI and Nino 3.4, (d) IOD and Nino 3.4, (e) IOD and SST, and (f) Nino and SST, with a lag of 120 months. The cross-correlation coefficients indicate significant coupling among these indices (Nino, IOD, and SST). All plots were generated using monthly temporal datasets. The figure was reproduced from Yadav et al. \cite{YADAV2023}}
    \label{fig_ccf_1999_2008}
\end{figure}

\begin{figure}[H]
\centering
\subfigure[]{\includegraphics[width=0.32\textwidth]{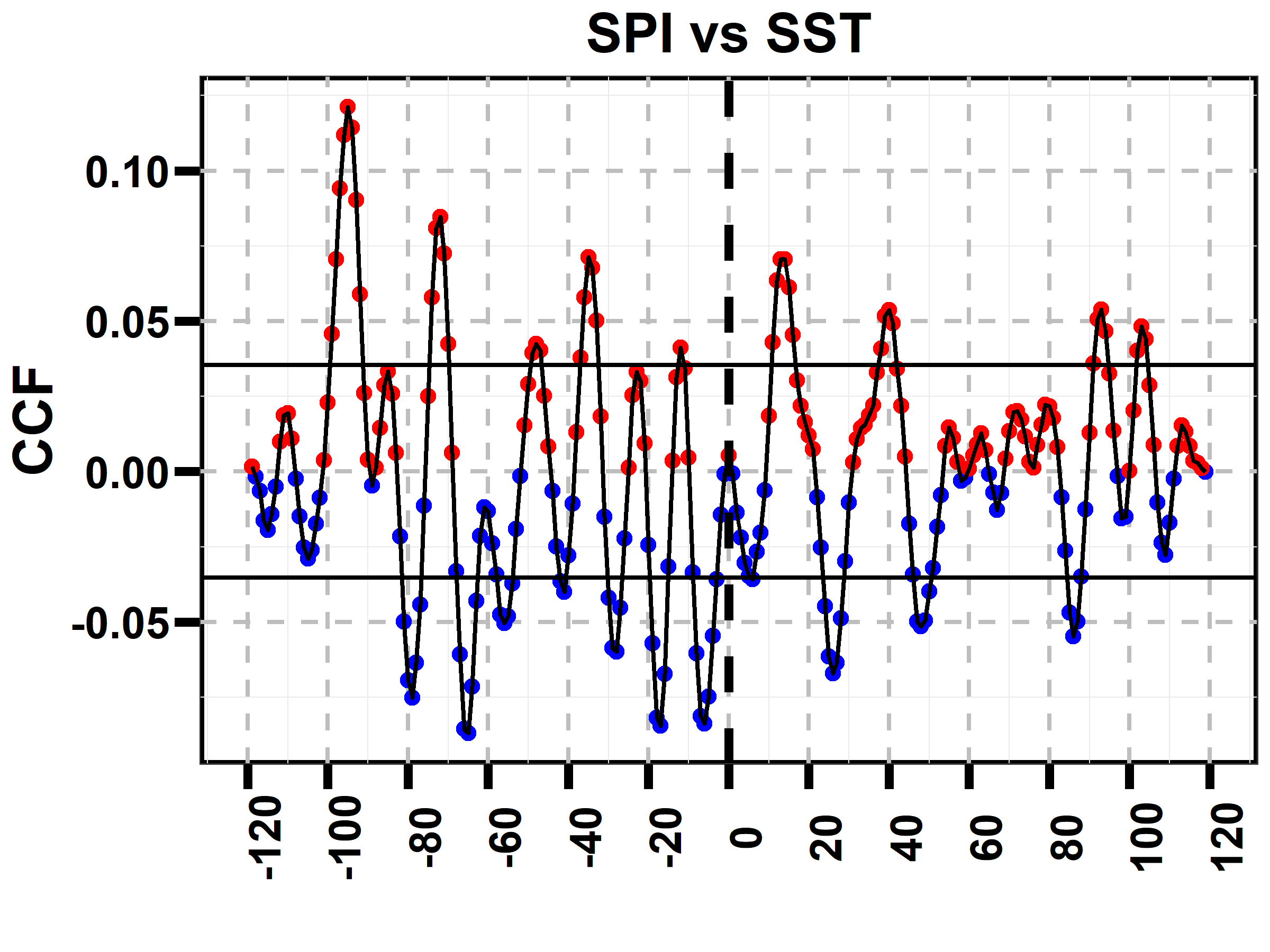}}
\subfigure[]{\includegraphics[width=0.32\textwidth]{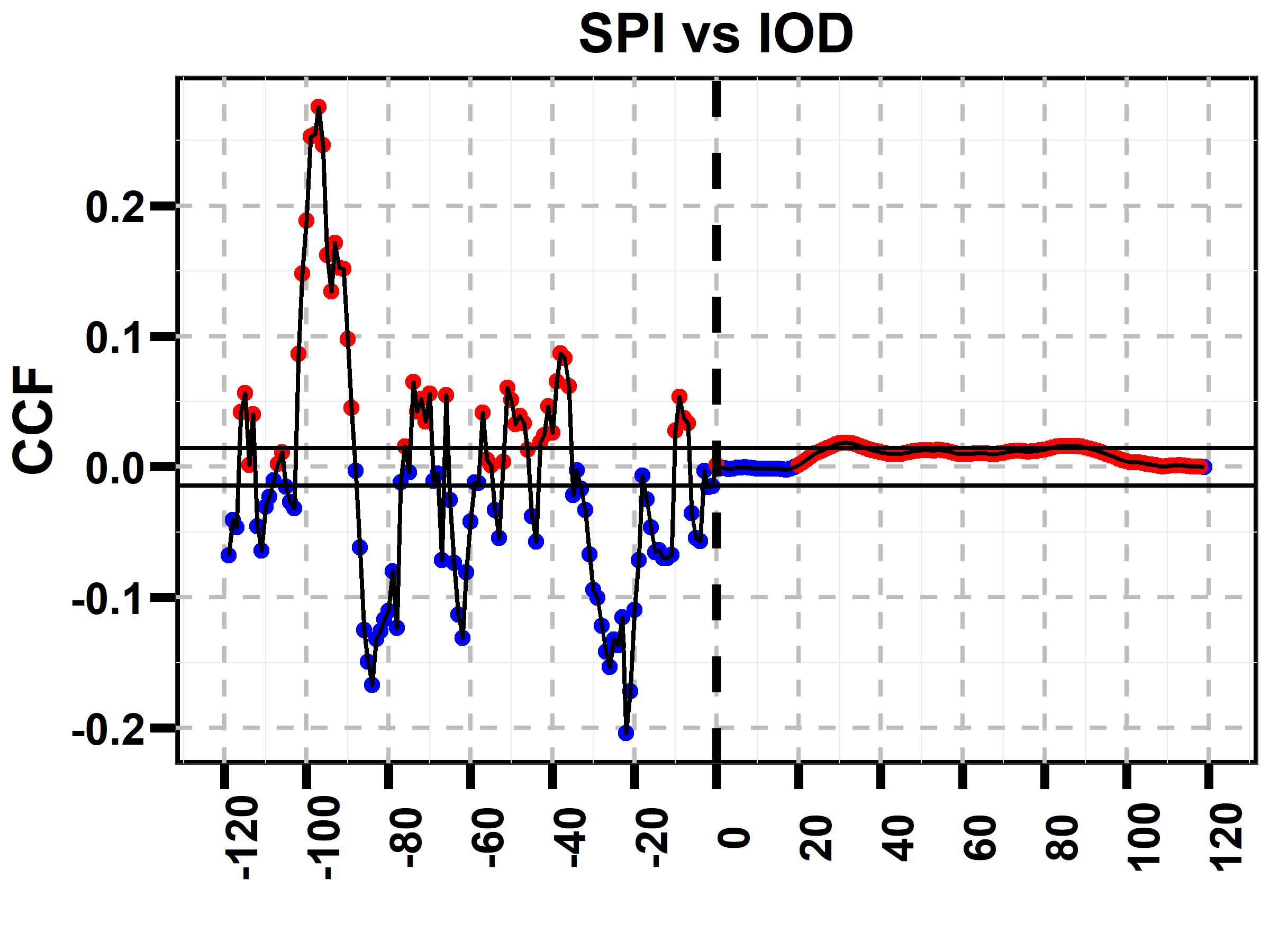}}
\subfigure[]{\includegraphics[width=0.32\textwidth]{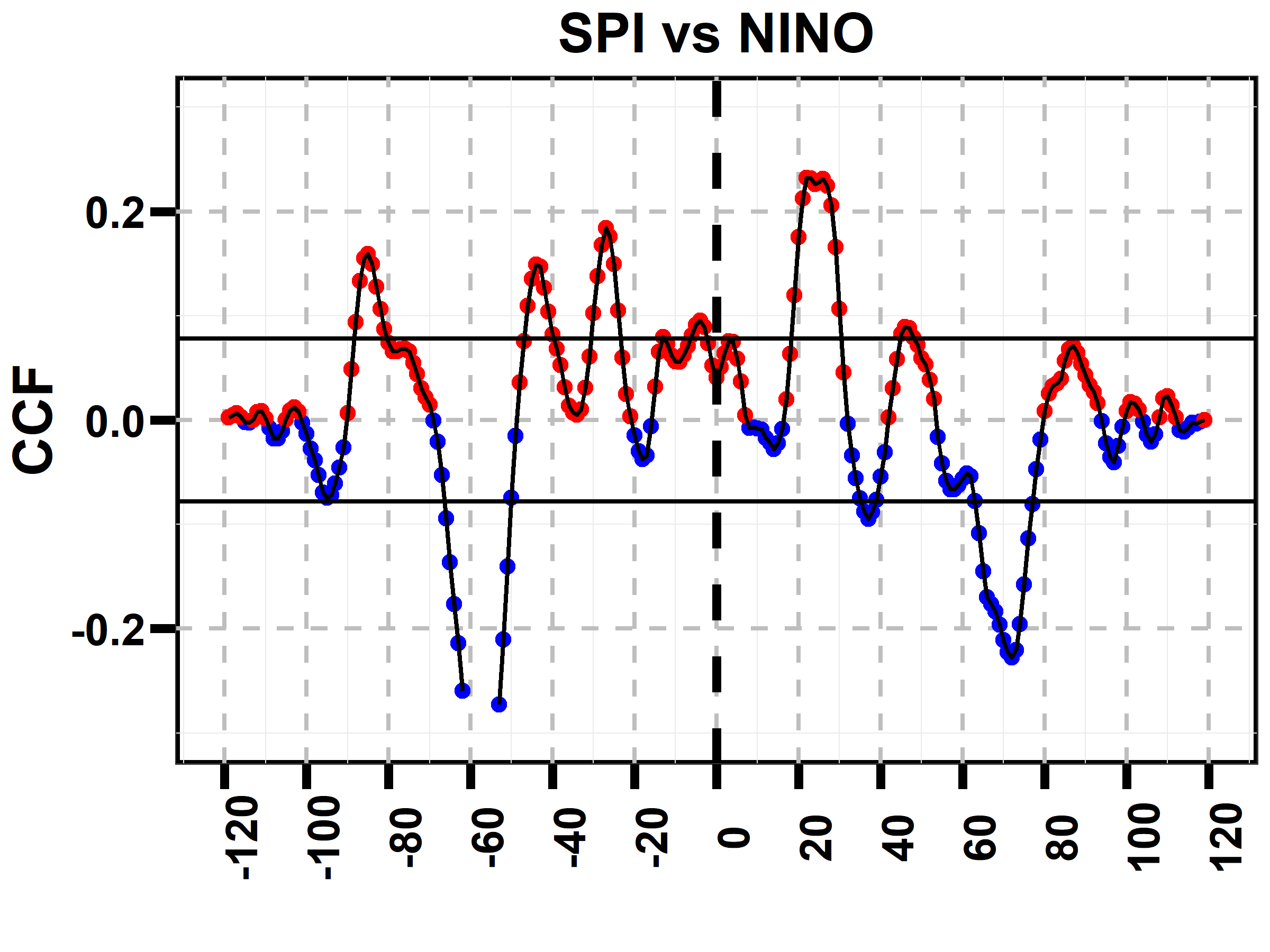}}
\subfigure[]{\includegraphics[width=0.32\textwidth]{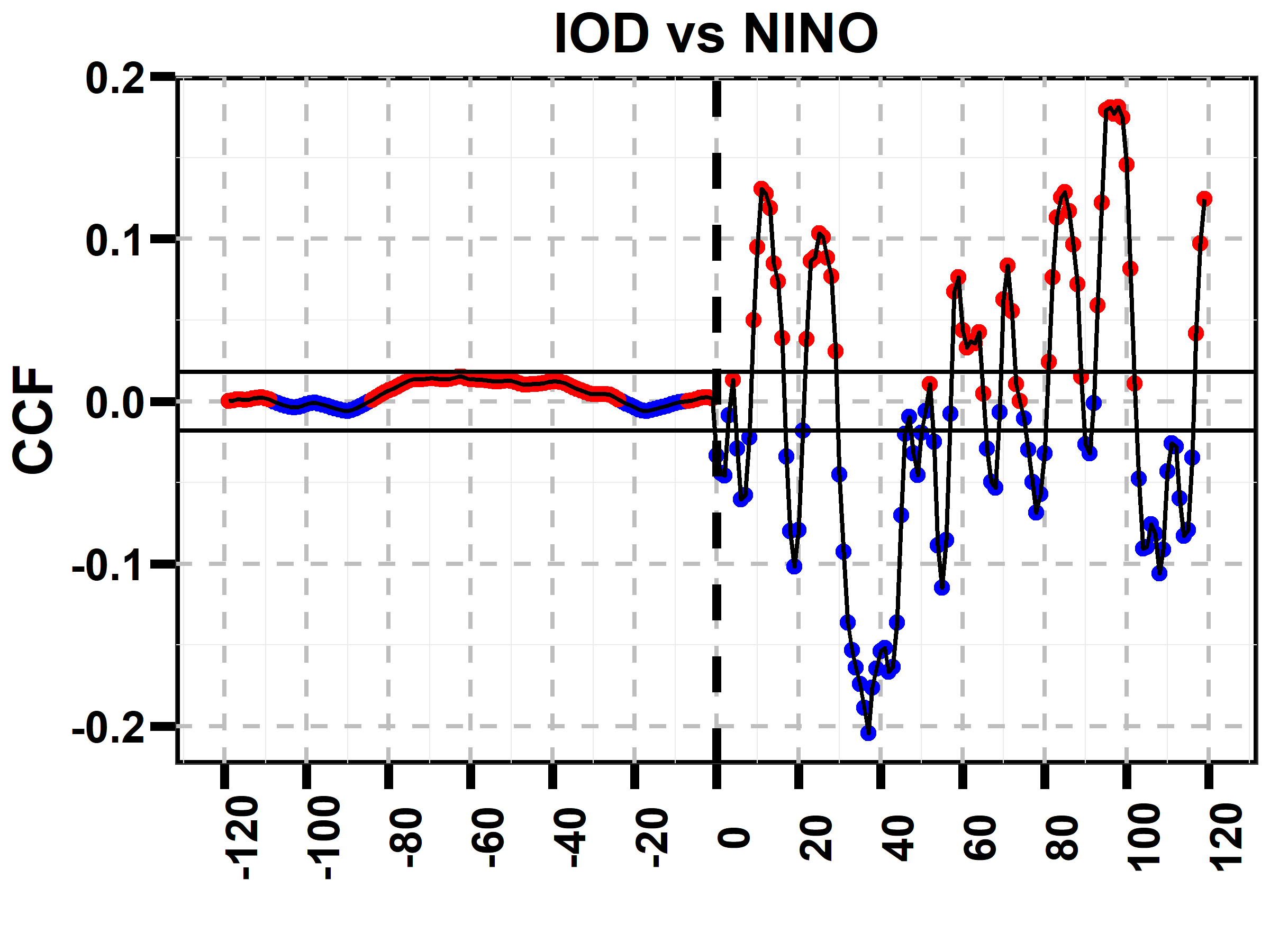}}
\subfigure[]{\includegraphics[width=0.32\textwidth]{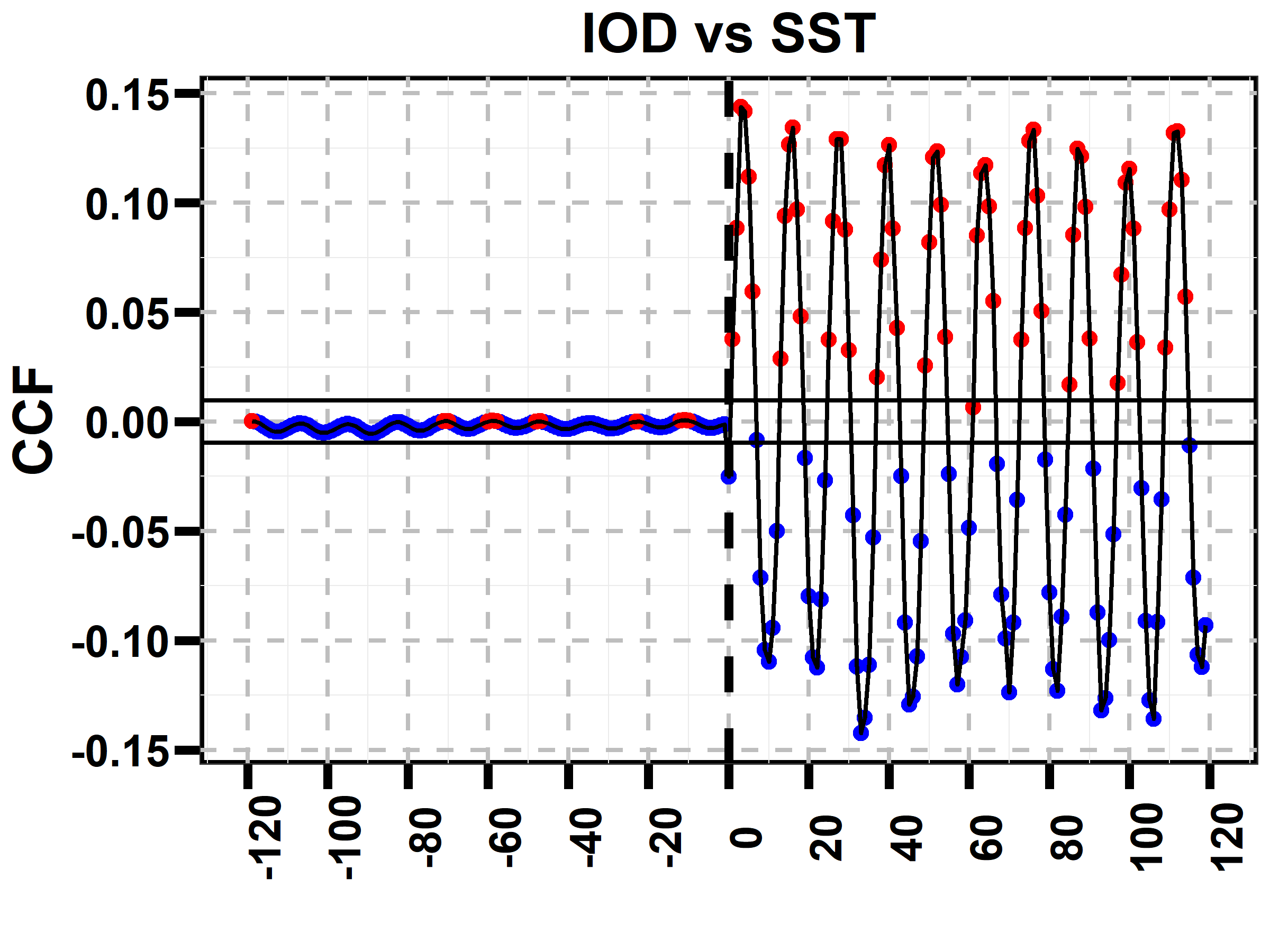}}
\subfigure[]{\includegraphics[width=0.32\textwidth]{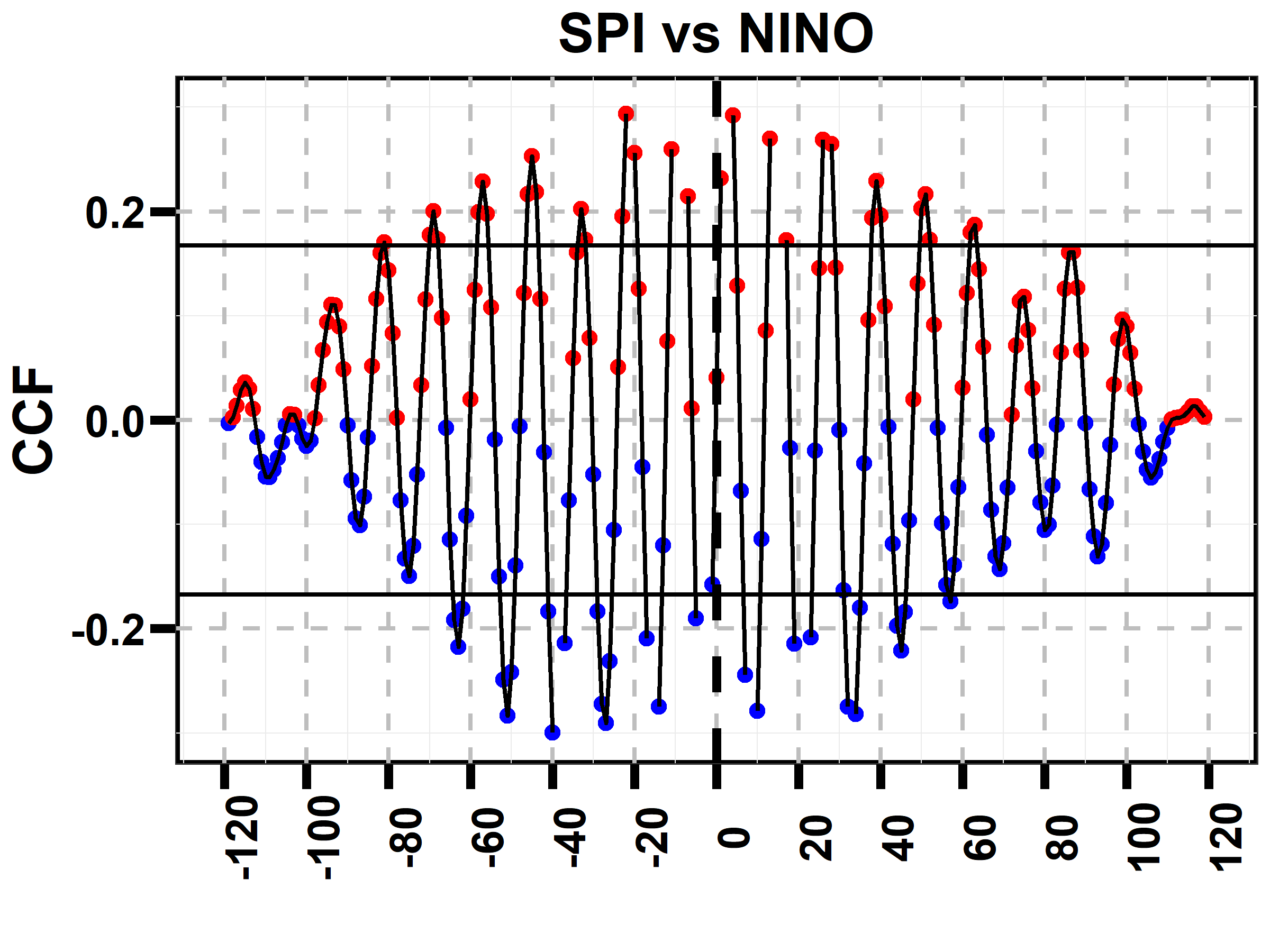}}
\caption{From 2009 to 2018, cross-correlation function (CCF) plots were generated for the following pairs: (a) SPI and SST, (b) SPI and IOD, (c) SPI and Nino 3.4, (d) IOD and Nino 3.4, (e) IOD and SST, and (f) Nino and SST, with a lag of 120 months. In Figure (a), despite a small CCF value, SPI and SST are observed to exhibit coupling. Figure (b) reveals a one-way cross-correlation between IOD and SPI, and vice versa, albeit statistically insignificant. Figure (c) highlights a significant coupling between Nino 3.4 and SPI. Conversely, Figure (d) indicates that while IOD influences Nino 3.4, the reverse is not observed. Figure (e) suggests a minor effect of IOD on SST, with no reciprocal influence observed. Notably, both SST and Nino 3.4 exhibit significant coupling. The substantial CCF coefficients indicate the significant influence of SST, Nino, and IOD on each other. Plots were generated using monthly temporal datasets. The figure was reproduced from Yadav et al. \cite{YADAV2023}}.
    \label{fig_ccf_2009_2018}
\end{figure}

\begin{figure}[H]
\centering
\begin{tabular}{cc}
    \includegraphics[width=.4\linewidth]{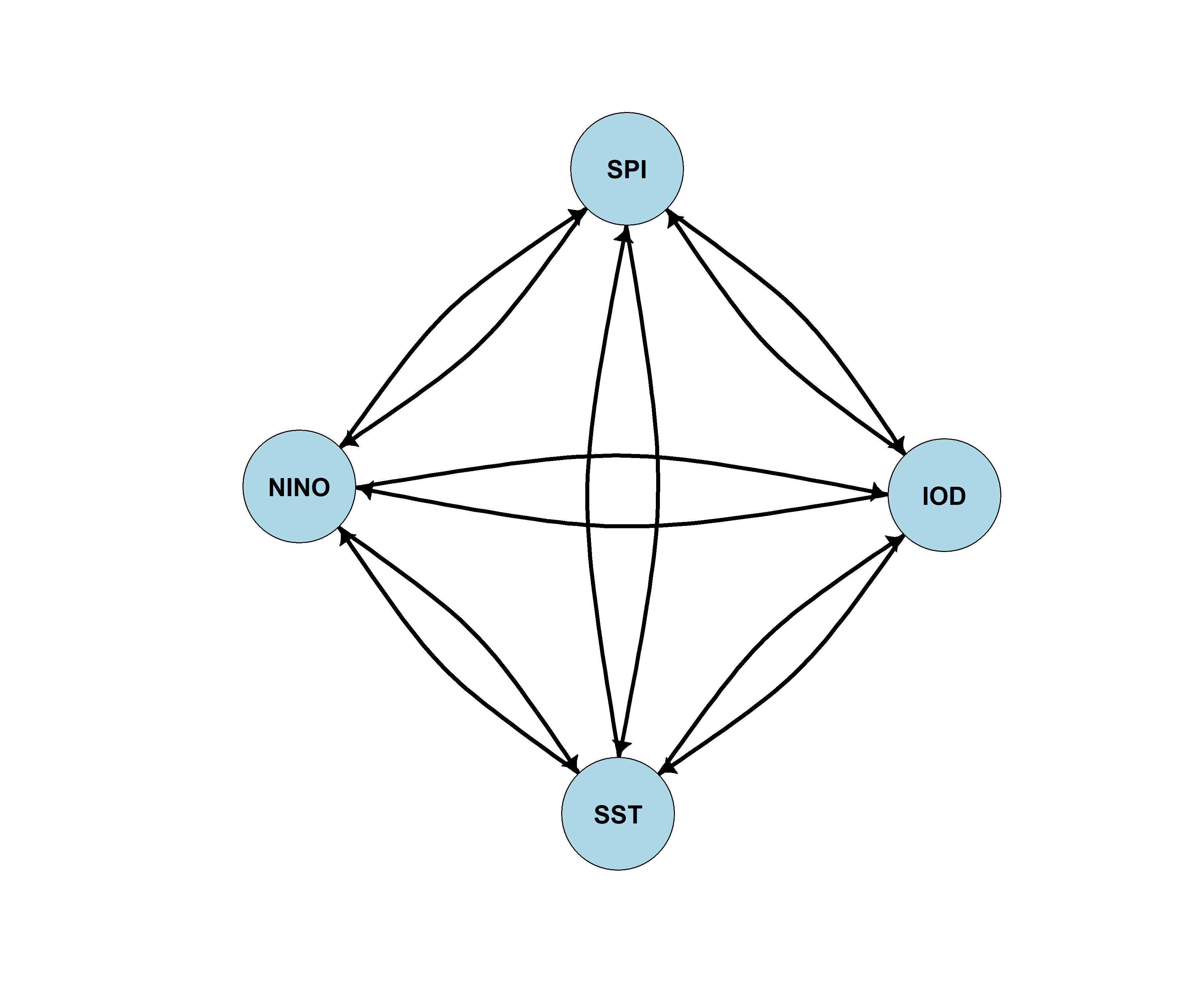}
 &  \includegraphics[width=.4\linewidth]{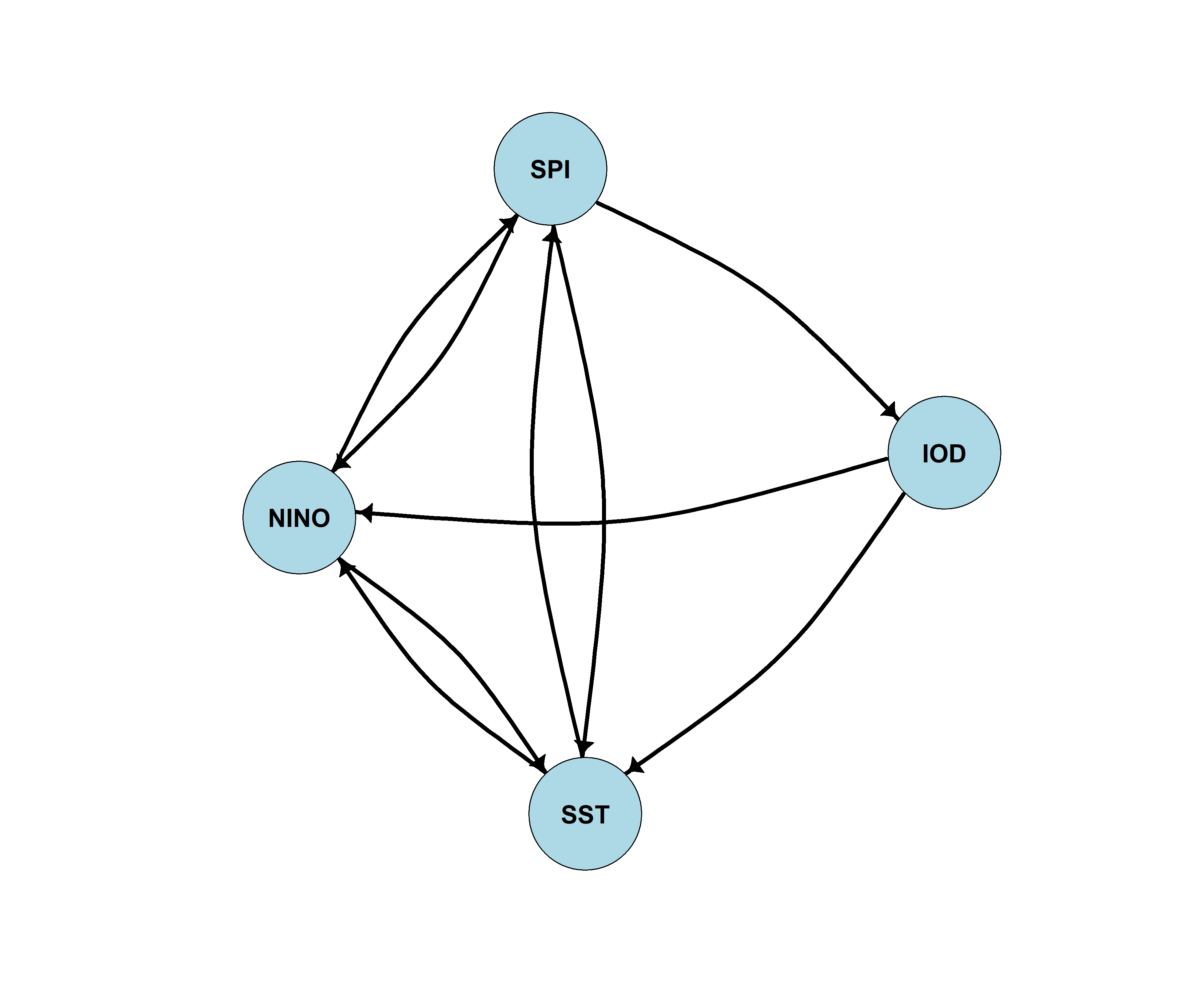}\\
 (a) & (b)\\
\end{tabular}
	\caption{The network analysis based on Cross-Correlation plots (\ref{fig_ccf_1999_2008}) and (\ref{fig_ccf_2009_2018}) for the periods (a) 1999 to 2008 and (b) 2009 to 2018 reveals distinct patterns. In Figure (a), spanning 1999 to 2008, IOD exhibits direct effects on SST, Nino 3.4, and SPI. However, in the subsequent decade (Figure (b), 2009 to 2018), IOD's direct impact on SPI diminishes. Nevertheless, both SST and Nino 3.4  continue to directly influence SPI in both periods, and vice versa. Consequently, IOD indirectly couples with SPI. The significant CCF coefficients indicate substantial mutual influences among these climate variables. The figure was reproduced from Yadav et al. \cite{YADAV2023}}
	\label{fig_network}
\end{figure}

\begin{table}[H]
\begin{center}
\begin{tabular}{l|cc|cc|cc|cc }
    \hline
\bf{Model}   & \multicolumn{2}{c|}{\bf{Type I}}
            & \multicolumn{2}{c|}{\bf{Type II}}
                    & \multicolumn{2}{c|}{\bf{Type III}}
                            & \multicolumn{2}{c}{\bf{Type IV}}                \\
\textbf{SPI}   &    \textbf{1m}  &    \textbf{12m}  &    \textbf{1m}  &    \textbf{12m}  &    \textbf{1m}  &    \textbf{12m}  &    \textbf{1m}  &    \textbf{12m}  \\
    \hline
\textbf{Model without }  &  \textbf{0.75 }  &   1.26  &   0.80  &   \textbf{0.49}  &   0.80  &  \textbf{0.51}   &   0.84  &   \textbf{0.38 } \\
\textbf{Spatial Correction}    &       &       &       &       &       &       &       &       \\\hline
\textbf{Model with }   &    7.21   &     \textbf{0.96}  &   6.40    &  \textbf{0.45}     &   8.66    & \textbf{0.46}      &   24.21    &    \textbf{0.37}   \\
\textbf{Spatial Correction }  &       &       &       &       &       &       &       &       \\ \hline
 \textbf{Model with LASSO }  &   \textbf{0.75}    &   0.77    &   0.80     &  \textbf{0.35 }    &     0.80 &     \textbf{0.34}  &    0.84   &    \textbf{0.34}   \\
 \textbf{and Spatial Correction}   &       &       &       &       &       &       &       &       \\ \hline

    \hline
\end{tabular}
\end{center}
\caption{The out-of-sample Root Mean Square Error (RMSE) values are calculated for all four types of models from December 2010 to November 2018, covering a period of 8 years. Here, the target variable is SPI (1 month and 12 months). A brief overview of the model types is as follows: Type I captures long-term memory solely using Fourier Series methods. Type II incorporates NINO 3.4 and IOD as covariates. Type III includes lag values of NINO 3.4 and IOD as covariates. Type IV incorporates NINO 3.4, IOD, and SST as covariates. It is observed that Type III and Type IV, considering LASSO and GP correction, yield the lowest (best) RMSE value of 0.34 for estimating SPI-12 months. The table was reproduced from Yadav et al. \cite{YADAV2023}}
\label{tab:RMSE}
\end{table}

\begin{figure}[H]
\centering
\includegraphics[width=0.6\linewidth]{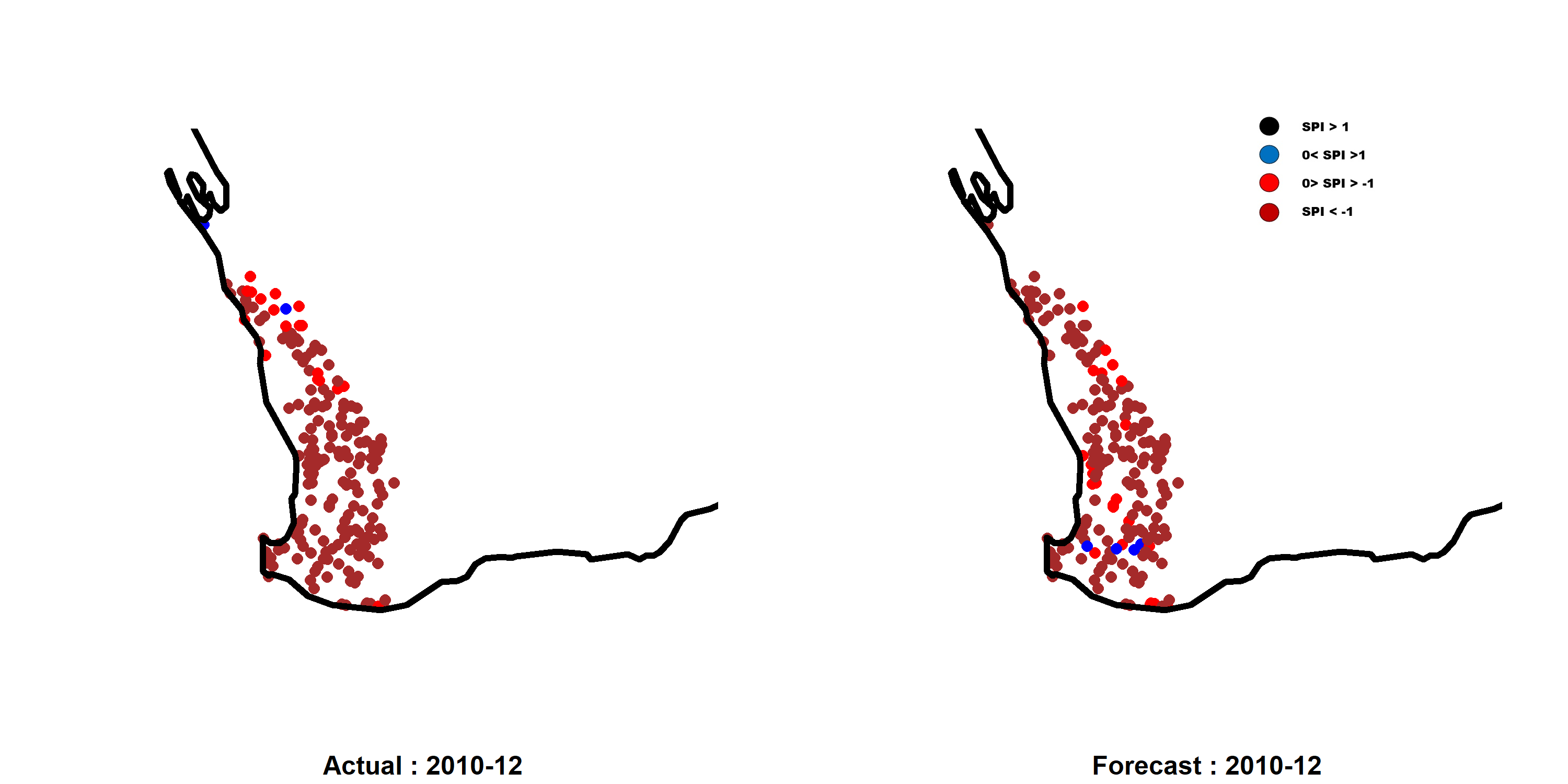}
\caption{Plot of the Standardized Precipitation Index (SPI) for December 2010 using training data ranging from June 1973 to November 2010. The SPI values are categorized into four color ranges: Black represents SPI > 1 (Extremely Wet), Blue represents 0 < SPI > 1 (Wet), Red represents 0 > SPI > -1 (Dry), and Brown represents SPI < -1 (Extremely dry). The figure was reproduced from Yadav et al. \cite{YADAV2023}}
\label{fig_dec_2010}
\end{figure}

\begin{figure}[H]
    \centering

    \subfigure[]{\includegraphics[width=0.32\textwidth]{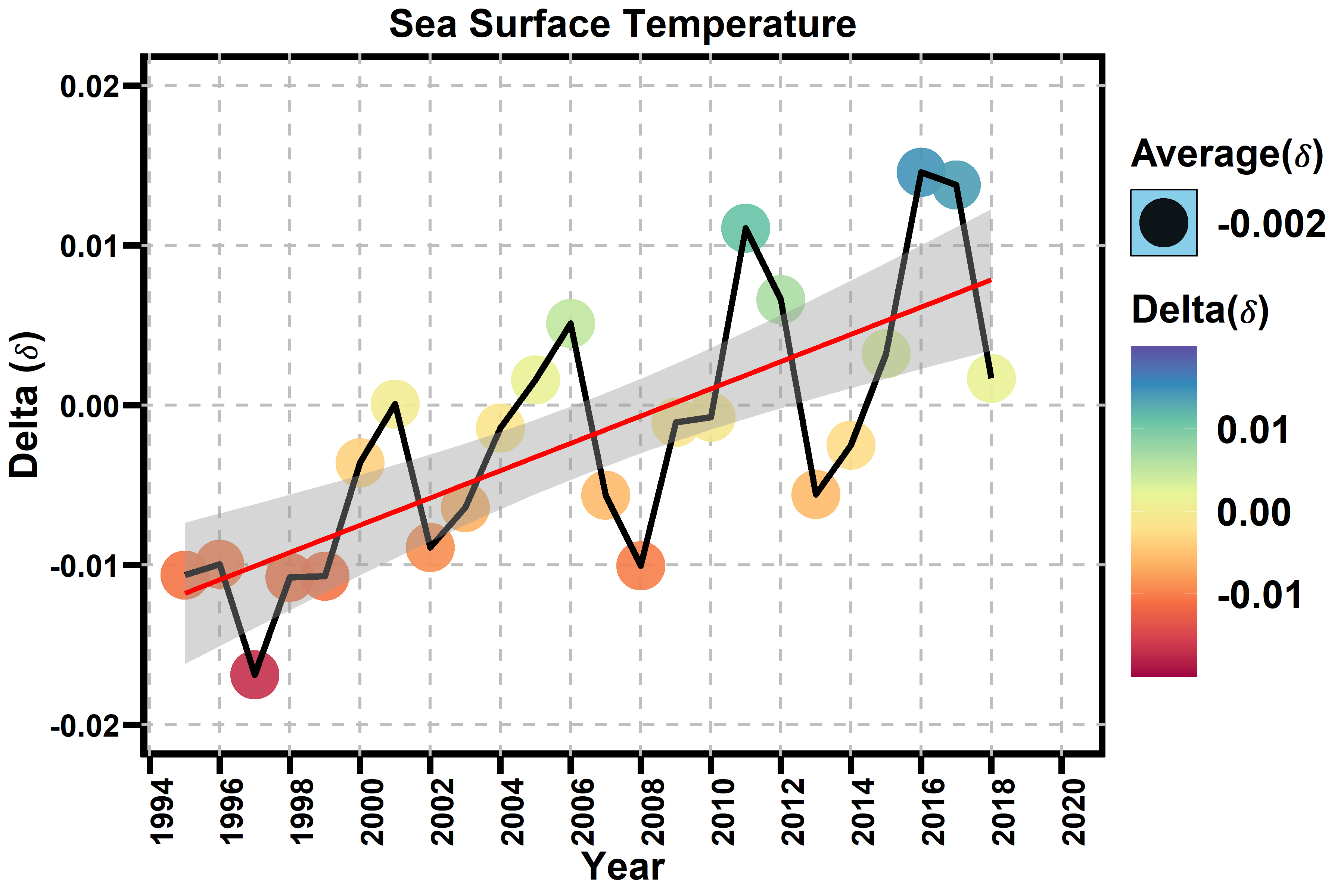}}
    \subfigure[]{\includegraphics[width=0.32\textwidth]{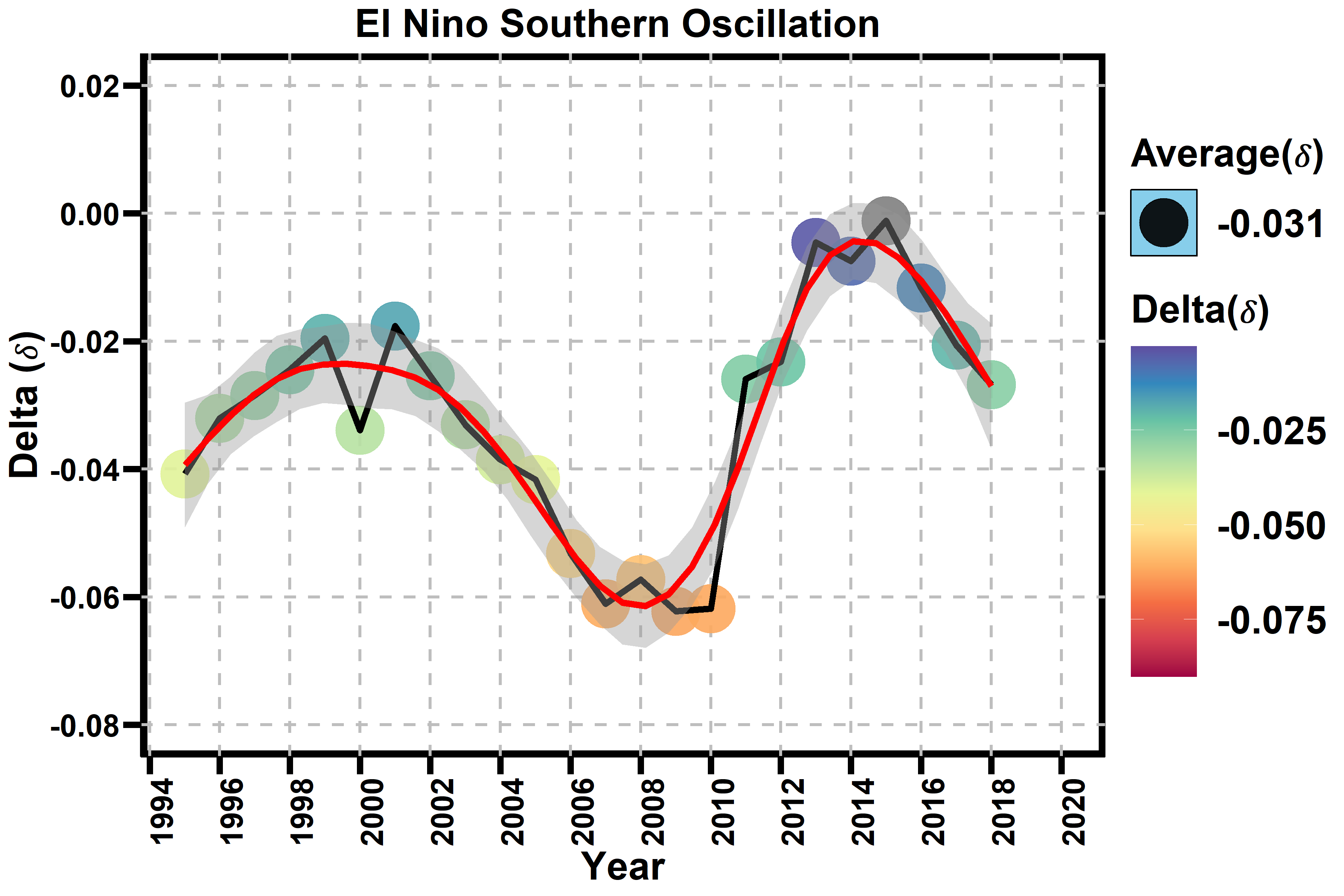}}
    \subfigure[]{\includegraphics[width=0.32\textwidth]{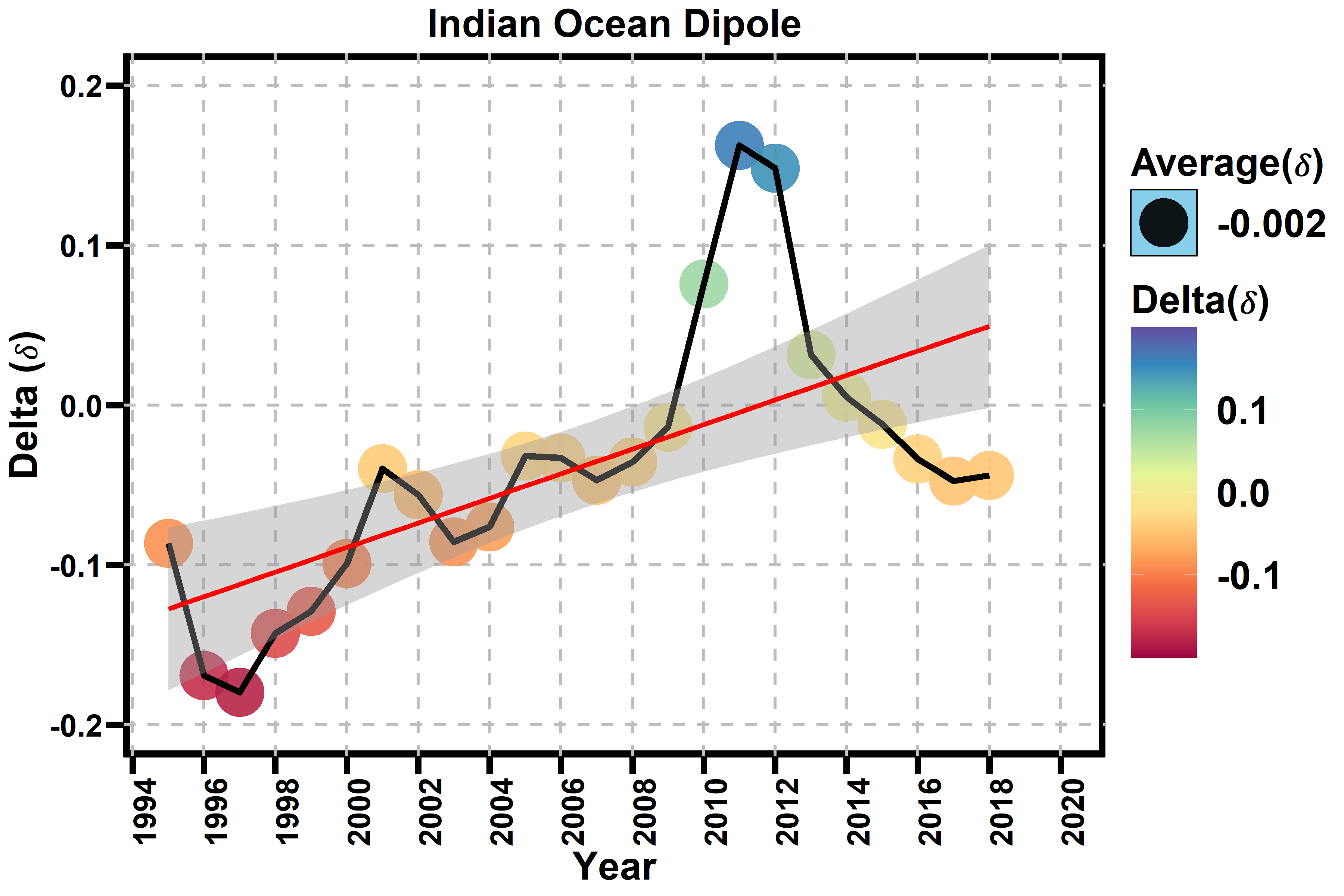}}
  
    \caption{Plots illustrating the average regression coefficient value ($\delta$) corresponding to NINO 3.4, SST, and IOD are displayed for the period from 1995 to 2018. A red smooth curve represents the three-year moving average of $\delta$. The analysis highlights that NINO 3.4 consistently exerts a significant negative impact on SPI. In contrast, SST exhibits a consistent negative correlation with SPI until 2004. However, between 2005 and 2014, the $\delta$ of SST fluctuates between negative and positive correlations. IOD demonstrates a negative correlation with SPI until 2009, followed by a positive range from 2009 to 2013, and then returns to a significantly negative correlation with SPI thereafter. The figure was reproduced from Yadav et al. \cite{YADAV2023}}
    
     \label{fig_ts_beta_sst_nino}
    \end{figure}

\subsection{North Atlantic Climate Instabilities}

Our second study incorporates three distinct datasets: daily mean Arctic Sea Ice Extent (SIE), daily mean Sea Surface Temperature (SST), and daily mean North Atlantic Oscillation (NAO) index. 

\subsubsection*{Data Description}
The NAO and SST datasets are sourced from the NOAA website \cite{NAO_Data_Source, SST_Data_Source}, while the SIE dataset is obtained from the National Snow and Ice Data Centre's website \cite{SIE_Data_Source}. These datasets cover various time periods, with NAO data available from 1950, SIE data from 1979, and SST data from 1982. Our analysis spans from January 1982 to September 2019, encompassing 38 years.

\subsubsection*{Results and Analyses}

Figure (\ref{fig_Long_memory_of_NAO}a) depicts the NAO plot spanning from 1979 to 2019. NAO, which signifies the difference in sea-level air pressure between the Icelandic Low and the Azores, demonstrates a stationary process with a mean of zero.
The time series plot confirms the mean-zero stationary nature of NAO. This observation was further validated through the Augmented Dickey-Fuller test, as evidenced by the small p-value. Additionally, the Auto-correlation function (ACF) analysis illustrated in Figure (\ref{fig_Long_memory_of_NAO}b) with a maximum lag of 5000 days (approximately 13 years) reveals the presence of long memory.

\begin{figure}[ht]
    \centering

    \subfigure[]{\includegraphics[width=0.32\textwidth]{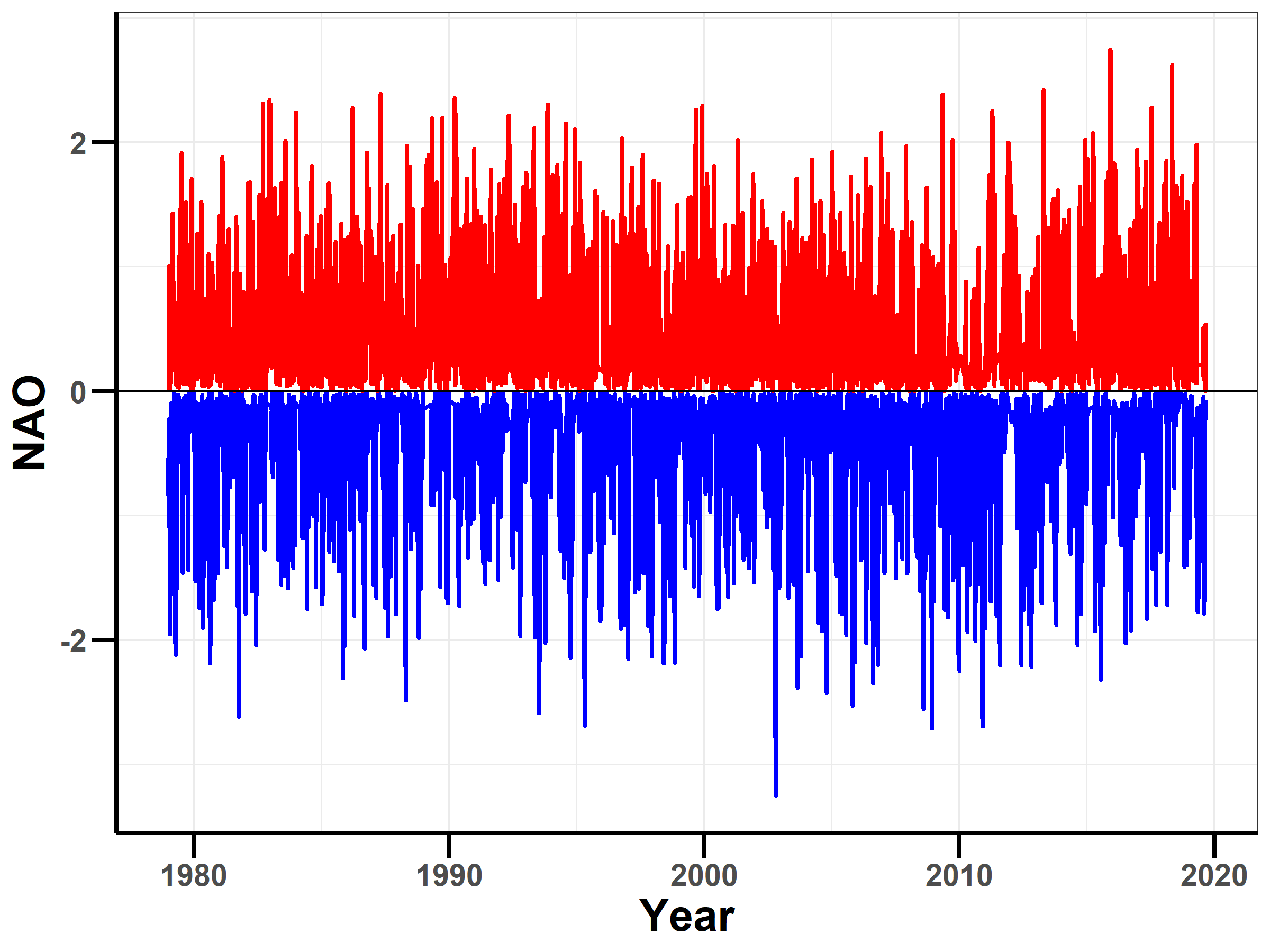}}
    \subfigure[]{\includegraphics[width=0.32\textwidth]{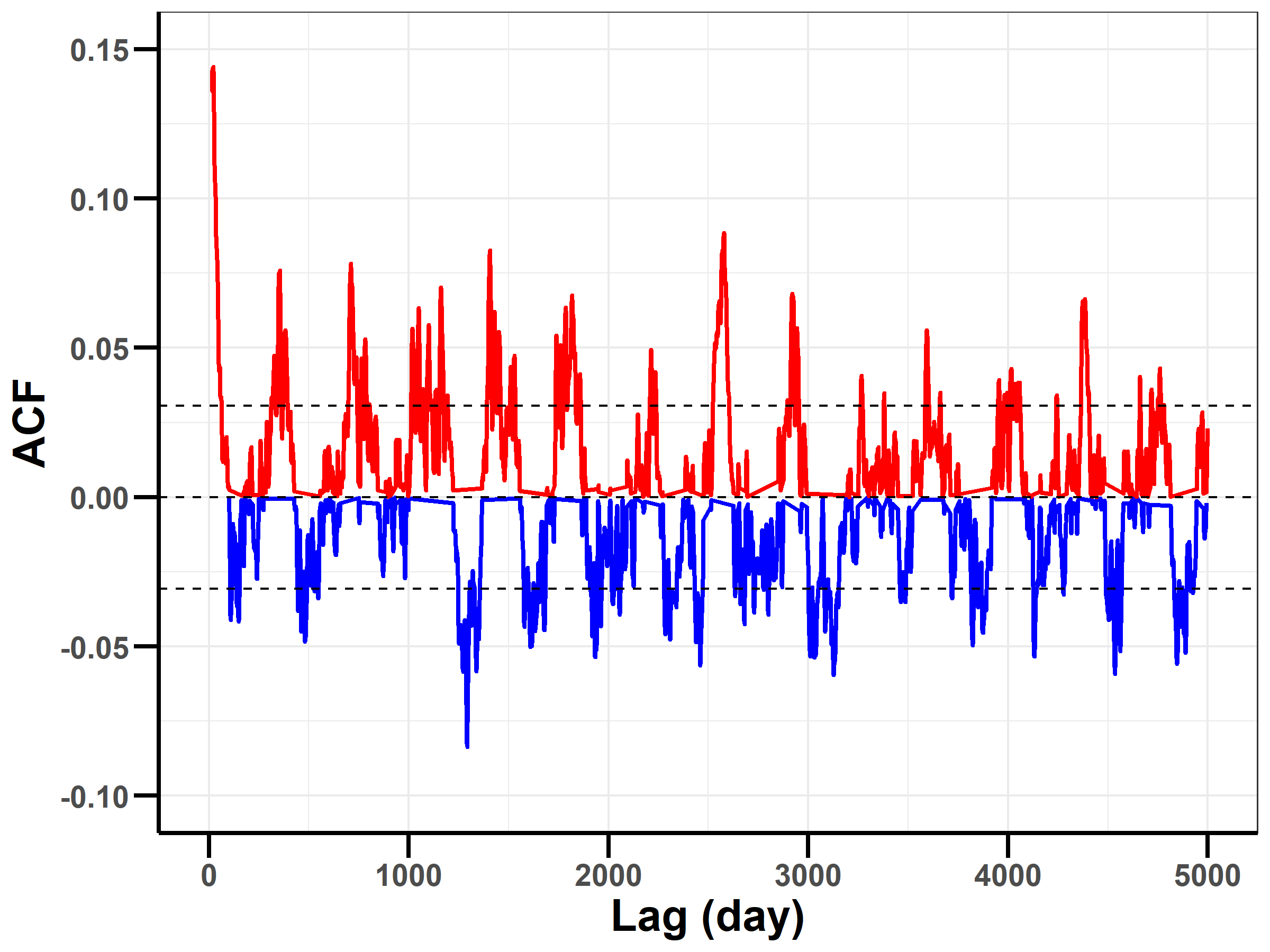}} 
    
    \caption{ (a) Plot of the NAO time series over the period of 1979–2019 (b) Autocorrelation plot of NAO with a maximum lag of 5000 days (almost 13 years), indicating the existence of long memory. The figure was reproduced from Yadav et al. \cite{yadav2023NAO}} 
    \label{fig_Long_memory_of_NAO}
\end{figure}


Figure(\ref{fig_ts_of_SIER_SSTR}) represents the residual for SIE and SST obtained from the best fit of Model (\ref{eqn_full_model}). It is observed that both SIER and SSTR  exhibit characteristics of zero-mean stationary processes, similar to the North Atlantic Oscillation (NAO). 

\begin{figure}[ht]
 \centering
\subfigure[]{\includegraphics[width=0.32\textwidth]{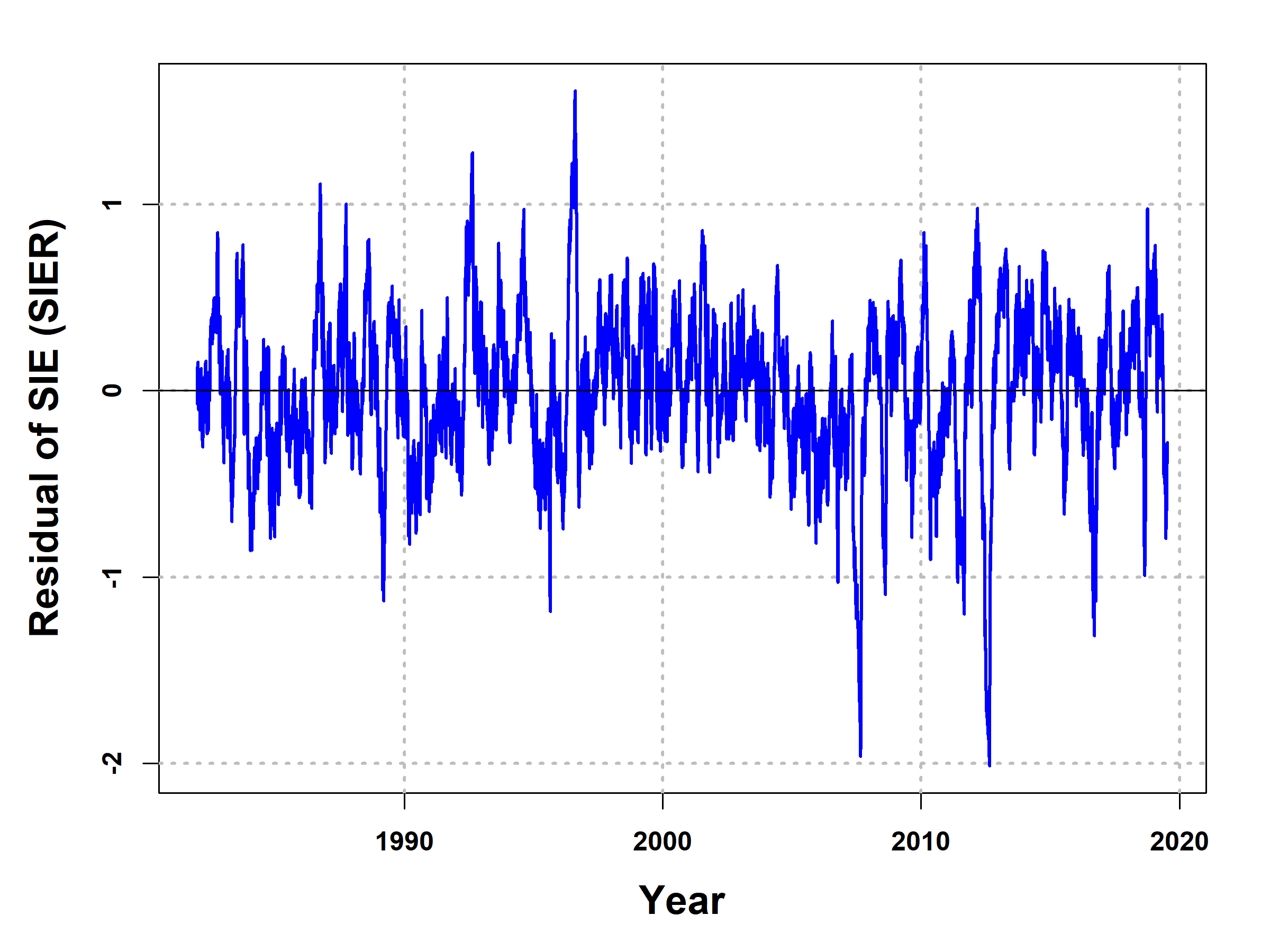}}
\subfigure[]{\includegraphics[width=0.32\textwidth]{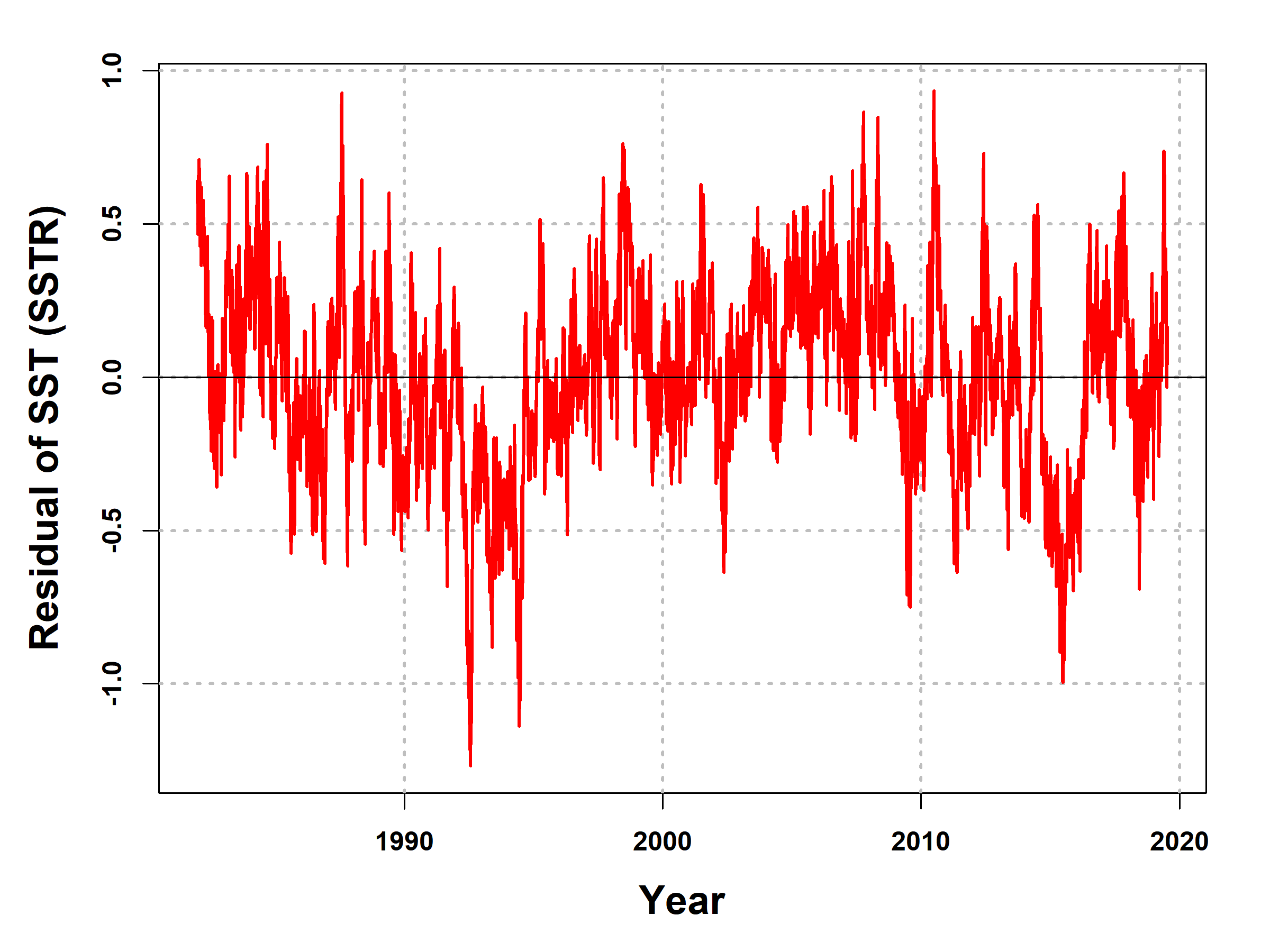}}
\caption{ Time series plots of (a) Residual of SIE (SIER) and (b) Residual of SST (SSTR). The figure was reproduced from Yadav et al. \cite{yadav2023NAO}}
  
 \label{fig_ts_of_SIER_SSTR}
\end{figure}

The Hurst exponent values detailed in Table (\ref{tab_Hurst_exp_SIER_SSTR}) collectively underscore the long memory inherent in residuals of SIE (SIER) and residual of SST (SSTR), with values surpassing 0.5. Additionally, the correlation matrix in Table (\ref{tab_NAO_SST_SIE_corr}) highlights robust correlations, particularly notable between SSTR and NAO, and between SIER and SSTR.

To explore the potential presence of a feedback loop between the SIER, SSTR, and NAO by employed the Granger causality test \ref{sec:Granger_Causal}. Our exploration into Granger causal models uncovers significant causal relationships among NAO, SIER, and SSTR. ANOVA F-tests reject null hypotheses, affirming Granger causality between these variables, as elaborated in Table (\ref{GC_models_result}). Furthermore, employing an Akaike information criterion-based model selection process helps identify optimal configurations for Model Equation \ref{sec: Modl_temporal}.

The synthesis of these findings highlights the existence of a reciprocal feedback loop involving NAO, SIER, and SSTR. Moving forward, we aim to demonstrate the affirmative nature of this loop. By emphasizing the skewness of NAO in Table (\ref{tab_skewness}), we provide insight into the bootstrap confidence intervals (C.I.) across various time intervals—daily, weekly, and monthly. In a scenario where NAO is stable, a skewness value of zero is anticipated. However, our results reveal a negatively skewed distribution, suggesting a statistically significant departure from stability.
 Notably, the skewness of NAO in Table (\ref{tab_skewness}) indicates a departure from stability, hinting at underlying instability within its dynamics.
In Figure (\ref{test_plot}), we present visualizations of projected and observed trajectories within the test dataset spanning 2010 to 2019. Our modeling approach demonstrates strong performance, as indicated by Root Mean Square Error (RMSE) values of 0.36 for the training set and 0.41 for the test set, underscoring its high generalization capability.

\begin{table}[ht]
\begin{center}
	 \setlength{\arrayrulewidth}{.8mm}
	\begin{tabular}{p{8cm}p{2cm}p{2cm}}
	\hline
	\bf{Hurst Exponent}  & \bf{SIER} & \bf{SSTR}\\
		\hline
		Simple R/S Hurst estimation   & 0.77 & 0.81\\
		Corrected R over S Hurst exponent   & 0.83 &0.89\\
		Empirical Hurst exponent   &0.82 &  0.90\\
		Corrected empirical Hurst exponent  &0.81 & 0.89\\
		Theoretical Hurst exponent   & 0.53 & 0.52\\ \hline
	\end{tabular}
		 \caption{The Hurst Exponent value of the SIER, and SSTR using different methods. The table was reproduced from Yadav et al. \cite{yadav2023NAO}}
		 \label{tab_Hurst_exp_SIER_SSTR}
\end{center}
\end{table}

\begin{table}[ht] 
	\centering
\setlength{\arrayrulewidth}{.8mm}
\begin{tabular}{p{2cm}p{2cm}p{3cm}p{4cm}}
	\hline
		 & \bf{NAO} &\bf{ SIER} &  \bf{SSTR} \\
		\hline
		\bf{NAO } &  1.000 & 0.016 (0.063) &  -0.133 (\small{$<2.2*10^{-16}$})  \\
		\bf{SIER} & & 1.000 & -0.173 (\small{$<2.2*10^{-16}$})  \\
	\bf{SSTR } &    &  & 1.000 \\ 
	\hline
	\end{tabular}
		\caption{ Over a span of 38 years, from January 1982 to September 2019, the correlation matrix for NAO, SIER, and SSTR is examined. The accompanying P-values enclosed in parentheses provide insights into the significance of the correlations. Notably, the correlation between SSTR and NAO is statistically significant. Similarly, a robust correlation is observed between SIER and SSTR. However, the correlation between NAO and SIER exhibits relatively weak significance. The table was reproduced from Yadav et al. \cite{yadav2023NAO}}
	\label{tab_NAO_SST_SIE_corr}
\end{table}

\begin{table}[H]
\begin{center}
	 \setlength{\arrayrulewidth}{.8mm}
	\begin{tabular}{p{6cm}p{2.5cm}p{3cm}}
	\hline
	\bf{GC Models} &\bf{F-value} & \bf{p-value}\\
        \hline      
        \bf{ SSTR + SIER $\rightarrow$ NAO  } & 2.31 & 0.0178 \\  
        \bf{  NAO + SIER $\rightarrow$ SSTR } & 5.546 &$2.16\times10^{-6}$ \\ 
        \bf{   NAO + SSTR $\rightarrow$ SIER } & 7.714 & $2.27\times10^{-10}$\\ 
         \hline  
	\end{tabular}
		 \caption{ Table of F-value and p-value of different combinations of Granger Causal Models. Small p-values indicate that there is a feedback loop among NAO, SIER, and SSTR. The table was reproduced from Yadav et al. \cite{yadav2023NAO}}
		 \label{GC_models_result}
\end{center}
\end{table}

\begin{table}[ht]
\begin{center}
	 \setlength{\arrayrulewidth}{.8mm}
	\begin{tabular}{p{3cm}p{3cm}p{4cm}}
	\hline
	\bf{Period} & \bf{Skewness} & \bf{C.I.} \\
		\hline
		\textbf{Daily} &  -.210 & [-0.242, -0.169]\\
	    \textbf{Weekly}  & -.213 & [-0.305, -0.107]\\
		\textbf{Monthly}  &   -.194 & [-0.368, -0.005] \\ \hline
	\end{tabular}
		 \caption{The presented table highlights the skewness of NAO, along with bootstrap-derived confidence intervals (C.I.), across daily, weekly, and monthly time spans. While an anticipated stable NAO would exhibit a skewness of zero, our observations indicate a negatively skewed distribution. This statistically significant result underscores the presence of instability in NAO. The table was reproduced from Yadav et al. \cite{yadav2023NAO}}
		 \label{tab_skewness}
\end{center}
\end{table}

\begin{figure}[H]
   \centering
    \includegraphics[width=0.49\linewidth]{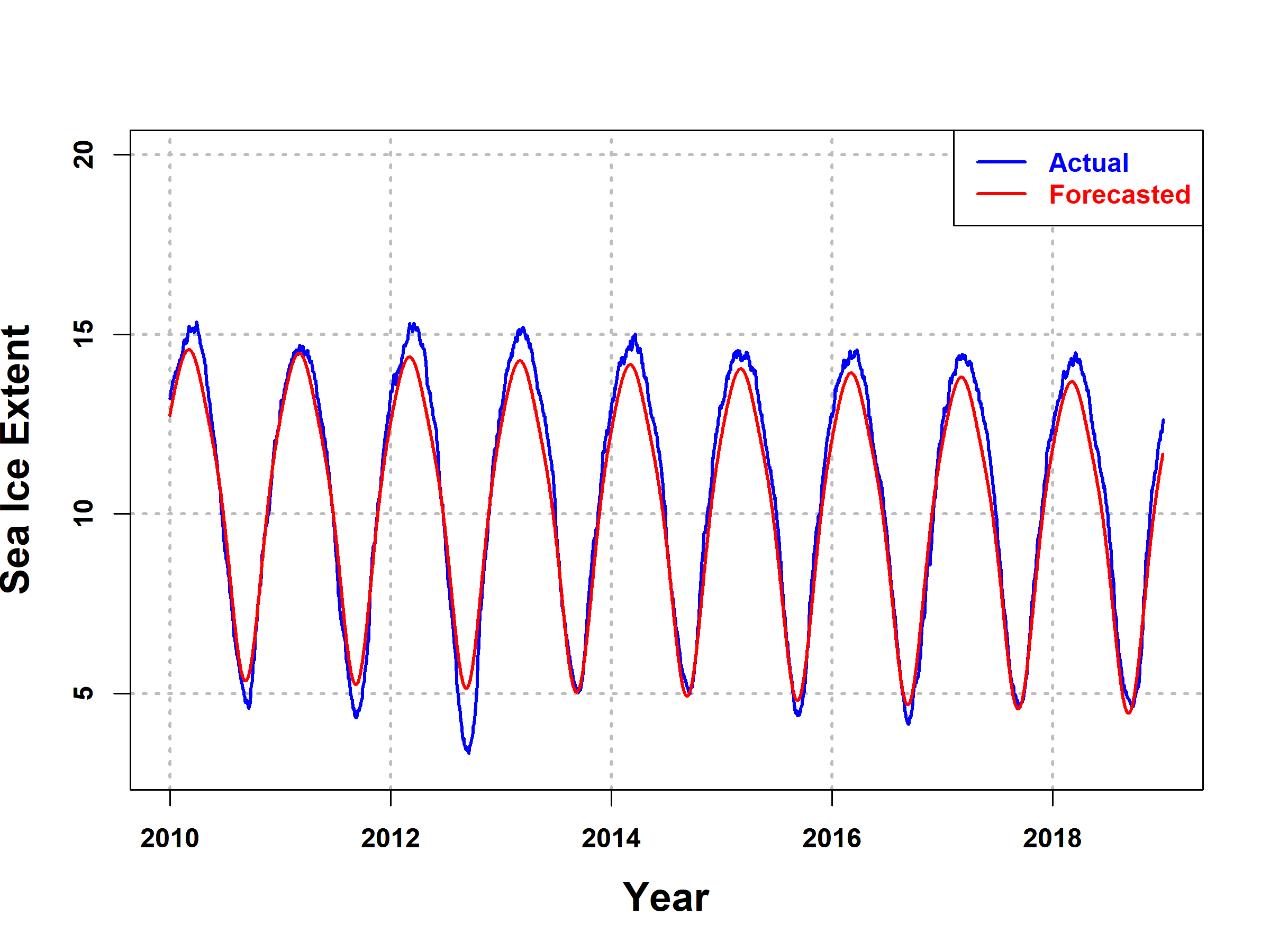}
    \caption{To evaluate the effectiveness of our proposed model a machine learning assessment was carried out. The training dataset spanned from 1979 to 2009, while the test dataset encompassed the years 2010 to 2019. The figure was reproduced from Yadav et al. \cite{yadav2023NAO}}
    \label{test_plot}
\end{figure}

Our analyses revealed crucial instability driven by positive feedback loops among melting SIE, rising SST, and the North NAO. Key findings included reciprocal Granger causality between SIE and SST, mutual Granger causality between SST and NAO, and an anti-correlation between SST and NAO. This anti-correlation suggested that increasing SST trends may lead to more negative NAO occurrences, a phenomenon supported by the negative skewness of the NAO index across various time scales.

The negative skewness of the NAO index, despite its mean-zero stationary nature, indicated impending critical instability, foreshadowing increased bouts of frigid climates in the North Atlantic region. This study contributed to predicting notable climate transformations by enhancing our understanding of critical instability within complex climate systems. Overall, through the application of statistical machine learning and interdisciplinary methods, as opposed to climate modelling, we tried to enrich the understanding of the dynamic interplay among crucial climate variables and its implications for the entire North Atlantic region.

\section{Outlook}

This review study contributes to untangling the complexity (transformations and critical instabilities) of climate systems. Through the application of statistical machine learning and interdisciplinary methods,  borrowed from physics, mathematics, statistics, etc. we tried to give a flavour of the alternate approaches to climate modeling. 

Climate change presents numerous challenges across various domains, including environmental, social, economic, and geopolitical aspects. Addressing these challenges requires coordinated efforts at local, national, regional, and global levels, including mitigation measures to reduce greenhouse gas emissions, adaptation strategies to build resilience, investments in clean energy and sustainable practices, policy interventions, technological innovations, education, and public awareness campaigns. We hope that these insights would pave the way for drafting policies for action against climate change and addressing its multifaceted challenges. Finally, 
 we accept our limitations as scientists, encapsulated by the famous quote:

\begin{quote}
{\it ``I used to think the top environmental problems were biodiversity loss, ecosystems collapse and climate change. I thought that with 30 years of good science we could address those problems. But I was wrong. The top environmental problems are selfishness, greed and apathy… and to deal with those we need a spiritual and cultural transformation and we, (Lawyers) and scientists, don't know how to do that.''} --
James Gustave Speth, US Advocate
\end{quote}

\section*{Acknowledgements}
The authors would like to thank K.S. Bakar for some of the collaborative work that has been included in this review article.


\end{document}